\documentclass[aps,prd,onecolumn,superscriptaddress,floats,amsfonts,amsmath,amssymb,linenumbers,12pt]{revtex4}

\pdfoutput=1
\usepackage{graphicx}
\usepackage{epstopdf}
\usepackage[active]{srcltx}
\usepackage{bm}
\usepackage[]{latexsym}

\usepackage{color}

\usepackage{caption2}

\usepackage{float}

\usepackage{lineno}

\begin{document}

\restylefloat{figure}

\title{Probing Neutrino Mass Hierarchy by Comparing the Charged-Current and Neutral-Current Interaction Rates of Supernova Neutrinos}


\author{Kwang-Chang Lai}
\email{kcl@mail.cgu.edu.tw}
\affiliation{Center for General Education, Chang Gung University, Kwei-Shan, Taoyuan, 333, Taiwan}
\affiliation{Leung Center for Cosmology and Particle Astrophysics (LeCosPA), National Taiwan University, Taipei, 106, Taiwan}

\author{Fei-Fan Lee}
\affiliation{Institute of Physics, National Chiao Tung University, Hsinchu, 300, Taiwan}

\author{Feng-Shiuh Lee}
\affiliation{Department of Electrophysics, National Chiao Tung University, Hsinchu, 300, Taiwan}

\author{Guey-Lin Lin}
\affiliation{Institute of Physics, National Chiao Tung University, Hsinchu, 300, Taiwan}
\affiliation{Leung Center for Cosmology and Particle Astrophysics (LeCosPA), National Taiwan University, Taipei, 106, Taiwan}

\author{Tsung-Che Liu}
\affiliation{Leung Center for Cosmology and Particle Astrophysics (LeCosPA), National Taiwan University, Taipei, 106, Taiwan}

\author{Yi Yang}
\affiliation{Department of Electrophysics, National Chiao Tung University, Hsinchu, 300, Taiwan}

\begin{abstract}

The neutrino mass hierarchy is one of the neutrino fundamental properties yet to be determined. We introduce a method to determine neutrino mass hierarchy by comparing the interaction rate of neutral current (NC) interactions, $\nu(\overline{\nu}) + p\rightarrow\nu(\overline{\nu}) + p$, and inverse beta decays (IBD), $\bar{\nu}_e + p\rightarrow n + e^+$, of supernova neutrinos in scintillation detectors. Neutrino flavor conversions inside the supernova are sensitive to neutrino mass hierarchy. Due to Mikheyev-Smirnov-Wolfenstein effects, the full swapping of $\bar{\nu}_e$ flux with the $\bar{\nu}_x$ ($x=\mu,~\tau$) one occurs in the inverted hierarchy, while such a swapping does not occur in the normal hierarchy. As a result, more high energy IBD events occur in the detector for the inverted hierarchy than the high energy IBD events in the normal hierarchy. By comparing IBD interaction rate with the mass hierarchy independent NC interaction rate, one can determine the neutrino mass hierarchy.

\vspace{3mm}


\end{abstract}

\maketitle

\section{Introduction}

With almost two decades of efforts, a considerable progress in constraining the neutrino mixing parameters has been achieved \cite{Maki,Pontecorvo}, based on various oscillation experiments with atmospheric, solar, and terrestrial neutrinos \cite{GonzalezGarcia:2007ib}. It is now well established that the flavor states $\nu_e$, $\nu_\mu$, and $\nu_\tau$ are superpositions of the vacuum mass eigenstates $\nu_1$, $\nu_2$, and $\nu_3$ \cite{Strumia:2006db} and the three mixing angles, $\theta_{12}$, $\theta_{23}$, and $\theta_{13}$,  and two mass-squared differences, $\Delta^2_{21}=m^2_2-m^2_1$ and $\Delta^2_{31}=m^2_3-m^2_1$ are well constrained. However, the neutrino mass hierarchy, i.e., the sign of mass squared difference $\Delta^2_{31}=m^2_3-m^2_1$, remains undetermined. Although various techniques have been proposed to resolve the neutrino mass hierarchy, this question remains open to date and represents an important challenge in particle physics. Recent efforts on resolving the neutrino mass hierarchy include works based on reactor neutrinos \cite{Petcov:2001sy,Ge:2012wj,Li:2013zyd,Capozzi:2013psa} different baseline experiments \cite{Ishitsuka:2005qi}, Earth matter effects on supernova (SN) neutrino signal \cite{Lunardini:2003eh,Dasgupta:2008my}, spectral swapping of SN neutrino flavors \cite{Duan:2007bt}, rise time of SN $\nu_e$ light curve \cite{Serpico:2011ir}, $\nu_e$ and $\bar{\nu}_e$ light curves on the early accretion phase \cite{Chiu:2013dya}, analysis of meteoritic SN material \cite{Mathews:2011jq}, and detection of atmospheric neutrinos in sea water or ice \cite{Winter:2013ema}.

Among various proposals to identify neutrino mass hierarchy, intensive efforts have been devoted to studying neutrinos from galactic SNe. As these neutrinos propagate outward, they can experience significant flavor transitions before arriving at the terrestrial detectors. The flavor conversions caused by the well-known Mikheyev-Smirnov-Wolfenstein (MSW) effect \cite{Wolfenstein:1978,Mikheev:1985} depends on the neutrino mass hierarchy. In addition, it has been suggested that, due to the large neutrino number density in the deep region of the core, coherent $\nu-\nu$ forward scatterings may lead to collective flavor conversion $\nu_e\bar{\nu}_e\leftrightarrow\nu_x\bar{\nu}_x$ ($x=\mu,~\tau$) over the entire energy range. This collective effect on the flavor transition of SN neutrinos depends crucially on the neutrino mass hierarchy and may also leaves imprints on the neutrino spectra.

Most of the methods that use SN neutrinos for determining the neutrino mass hierarchy are based on the interactions of these neutrinos with atomic nuclei and free protons. The major interaction channel for neutrino detection is the inverse beta decay (IBD), $\bar{\nu}_e+p\rightarrow n+e^+$. While the liquid scintillation detector is sensitive to $\bar{\nu}_e$, the liquid argon detector has a good sensitivity to $\nu_e$ via charged-current interactions $^{40}{\rm Ar}+\nu_e\rightarrow^{40}{\rm K}^*+e^-$. While $\nu_e$ and $\bar{\nu}_e$ spectra have been intensively studied for understanding the physics of SN, the potential of utilizing other species of SN neutrinos as the tool is much less discussed. The detection of other species of SN neutrinos was proposed by measuring the neutral-current (NC) interactions, $\nu+p\rightarrow\nu+p$ \cite{Beacom:2002hs,Dasgupta:2011wg}. It has been shown that, under fairly optimistic assumption about the detector and SN neutrino flux, signals due to $\nu p$ elastic scatterings at scintillation detectors allow one to reconstruct $\nu_x$ spectrum and measure total neutrino energy to a precision $\leq10\%$ \cite{Dasgupta:2011wg}. 

In this work, we introduce an approach to determine the neutrino mass hierarchy, based on the interactions of SN neutrinos with free protons in scintillation detectors. The method is to explore how the inverse beta decays and neutral current interactions in single scintillation detector are related for different neutrino mass hierarchies. We provide calculations with contemporary inputs of several detectors: the presently-running Borexino \cite{Bellini:2009jr,Cadonati:2000kq,Miramonti:2003hw,Bellini:2010hy,Arpesella:2008mt} and KamLAND \cite{Araki:2005qa,Yoshida:2010}, the near-term SNO+ \cite{Tolich:2009,Maneira:2010,O'Keeffe:2011xi}, and the much larger proposed JUNO \cite{An:2015jdp,Adam} and LENA \cite{Wurm:2010ny,Wurm:2011zn}. 

The paper is organized as follows. In Sec. II, we review the supernova neutrino fluence. The flavor transitions of SN neutrinos are discussed as they propagate outward from the deep inside the supernova and traverse the Earth medium to reach the detector. In Sec. III, we define interaction spectra of inverse beta decay and neutral current and illustrate how to derive the interaction spectra from observations of scintillation detectors. Then, in Sec. IV, we present our calculations for considered detectors with statistical uncertainties addressed. In Sec. V, We extend our approach to the time-dependent SN neutrino flux assuming a scenario of the time evolution of SN neutrino luminosities.  We summarize and conclude in Sec. VI. 

\section{Supernova Neutrino Fluence}

\subsection{Primary Neutrino Fluence}

A SN emits a total energy ${\mathcal E}\approx10^{53}$ erg over a burst $\Delta t\approx10{\rm s}$ in neutrinos of all six flavors. The neutrino flavors $\nu_\mu$, $\nu_\tau$ and their antiparticles have similar interactions and thus similar average energies and fluences. Therefore, the total energy is divided as ${\mathcal E}={\mathcal E_{\nu_e}}+{\mathcal E_{\bar{\nu}_e}}+4{\mathcal E_{\nu_x}}$. In general, equipartition of energies among the primary neutrino flavors is expected in typical SN simulations, ${\mathcal E_{\nu_e}}\approx{\mathcal E_{\bar{\nu}_e}}\approx{\mathcal E_{\nu_x}}$ and also ${\mathcal L_{\nu_e}}\approx{\mathcal L_{\bar{\nu}_e}}\approx{\mathcal L_{\nu_x}}$ for luminosities, which is assumed in our calculation. The primary SN neutrino energy spectrum is typically not purely thermal. We adopt a Keil parametrization \cite{Keil:2002in} for the neutrino fluence
\begin{equation}
F^0_\alpha(E)=\frac{\Phi_\alpha}{<E_\alpha>}\frac{(1+\eta_\alpha)^{(1+\eta_\alpha)}}{\Gamma(1+\eta_\alpha)}\left(\frac{E}{<E_\alpha>}\right)^{\eta_\alpha}\exp\left[-(\eta_\alpha+1)\frac{E}{<E_\alpha>}\right],
\end{equation}
where $\Phi_\alpha={\mathcal E}_{\alpha}/<E_\alpha>$ is the time-integrated flux, $<E_\alpha>$ is the average neutrino energy, and $\eta_\alpha$ denotes the pinching of the spectrum.
If flavor conversions do not occur during the propagations of neutrinos from the SN core to the Earth, a SN at a distance $d$  thus yields a neutrino fluence
\begin{equation}
F_\alpha=\frac{F^0_\alpha}{4\pi d^2}=\frac{2.35\times 10^{13}}{\rm cm^2 MeV}\frac{{\mathcal E}_\alpha}{d^2}\frac{E^3}{<E_\alpha>^5}\exp\left(-\frac{4E}{<E_\alpha>}\right), \label{fluence}
\end{equation}
with ${\mathcal E}_\alpha$ in units of $10^{52}$ ${\rm erg}$, $d$ in 10 ${\rm kpc}$, and energies in ${\rm MeV}$.
For the numerical evaluations, we take a representative supernova at the Galactic center region with $d=10 ~{\rm kpc}$, and a total energy output of ${\mathcal E}=3\times10^{53}~{\rm erg}$, i.e., ${\mathcal E}_\alpha=5\times10^{52}~{\rm erg}$ for each of the 6 flavors. Further, we choose $<E_{\nu_e}>=12~{\rm MeV}$, $<E_{\bar{\nu}_e}>=15~{\rm MeV}$, and $<E_{\nu_x}>=18~{\rm MeV}$.
\label{fluence}
\subsection{Neutrino Flavor Transition inside SN}

Our knowledge on the neutrino flavor transition in a core-collapse SN suggests that the flavor conversion can be induced by the collective neutrino oscillation \cite{Dasgupta:2009mg,Dasgupta:2010ae,EstebanPretel:2007ec,Raffelt:2007cb,Hannestad:2006nj,Dasgupta:2007ws,Choubey:2010up,Duan:2005cp,Mirizzi:2010uz} (see \cite{Duan:2010bg} for a review)  and MSW effect. The collective oscillation results from the coherent $\nu-\nu$ forward scatterings in the deep region of the core where neutrino densities are large and may lead to collective pair conversion $\nu_e\bar{\nu}_e\leftrightarrow\nu_x\bar{\nu}_x$ ($x=\mu,~\tau$) over the entire energy range even with extremely small neutrino mixing angles. The MSW effect, instead, arise from neutrino interaction with ordinary stellar medium. In a typical SN, the collective flavor conversions would take place near $r\sim10^3~{\rm km}$ while those of MSW type would take place at  $r\sim10^4-10^5~{\rm km}$. As the collective and MSW effects are widely separated, they can be considered to be independent of each other. 

Recently, substantial progress has been made on the studies of neutrino collective flavor conversions. The most prominent feature arising from collective neutrino oscillations is the spectral swap/split. Analytical and numerical works have shown \cite{Dasgupta:2009mg,Raffelt:2010zza} that collective effects not only depend on neutrino parameters, SN environments, and primary neutrino spectra but also subject to the oscillation modes and simulation approaches one has chosen. However, recent study based on multi-angle analysis of SN neutrinos \cite{Sarikas:2011am,Chakraborty:2011nf} found that the seemingly dominant collective effects may be suppressed by the dense matter during the accretion phase following the core bounce \cite{EstebanPretel:2008ni}. 

Unlike the status of MSW effects, consensus on collective flavor transitions has not yet been reached. To avoid digression to diverse scenarios of the collective effect, we assume that MSW effect dominates the flavor conversions when SN neutrinos propagate outwards. We will also neglect the complicated effects during the cooling phase since we are mainly interested in the time-integrated flux.

\subsection{Neutrino Fluence on Earth}

As neutrinos propagate outwards from deep inside SN and finally reaches the Earth, their flavor contents are modified by the MSW effect. Let us denote the survival probability for $\nu_e(\bar{\nu_e})$ after the MSW effect as $P(\bar{P})$. Then, the fluxes of $\nu_e$ and $\bar{\nu}_e$ arriving at the detector can be written as:
\begin{eqnarray} 
F_e            & = & P F^0_e + (1-P) F^0_x,  \\
F_{\bar{e}} & = & \bar{P} F^0_{\bar{e}} + (1-\bar{P}) F^0_{\bar{x}}, 
\end{eqnarray}
with
\begin{eqnarray} 
P            & = & P_{1e} P_H P_L + P_{2e}(P_H-P_H P_L) + P_{3e}(1-P_H), \\
\bar{P} & = & \bar{P}_{1e} (1-\bar{P}_L) + \bar{P}_{2e} \bar{P}_L,
\end{eqnarray}
for the normal hierarchy, and
\begin{eqnarray} 
P            & = & P_{1e} P_L + P_{2e}(1 - P_L), \\
\bar{P} & = & \bar{P}_{1e} \bar{P}_H (1-\bar{P}_L) + \bar{P}_{2e} \bar{P}_H \bar{P}_L + \bar{P}_{3e}(1-\bar{P}_H),
\end{eqnarray}
for the inverted hierarchy. Here, $P_H$ ($\bar{P}_H$) and $P_L$ ($\bar{P}_L$) are the crossing probabilities for the neutrino (antineutrino) eigenstates at higher and lower resonances, respectively. $P_{ie}$ ($\bar{P}_{ie}$) is the probability that a mass eigenstate $\nu_i$ ($\bar{\nu}_i$) is observed as a $\nu_e$ ($\bar{\nu}_e$) since neutrinos arrive at the Earth as mass eigenstates.
With the recent determination of the relatively large $\theta_{13}$, the flavor crossings are adiabatic and the vanishing crossing probabilities $P_H\simeq\bar{P}_H\simeq P_L\simeq\bar{P}_L\simeq0$ can be adopted \cite{Dighe:1999bi}. The fluxes then become
\begin{eqnarray} 
F_e            & = &  F^0_x,   \label{eNH}  \\
F_{\bar{e}} & = &  (1-\bar{P}_{2e}) F^0_{\bar{e}} + \bar{P}_{2e} F^0_{\bar{x}}, \label{ebarNH}
\end{eqnarray}
for the normal hierarchy, and
\begin{eqnarray} 
F_e            & = & P_{2e} F^0_e + (1-P_{2e}) F^0_x,  \label{eIH} \\
F_{\bar{e}} & = & F^0_{\bar{x}}, \label{ebarIH}
\end{eqnarray}
for the inverted hierarchy \cite{Dighe:1999bi}. Here the probability $P_{2e}$ is usually written as $P_{2e}=\sin^2\theta_{12}+f_{\rm reg}$, with $f_{\rm reg}$ the regeneration factor due to the Earth matter effect \cite{deHolanda:2004fd}:
\begin{equation}
f_{\rm reg}=\frac{2E\sin^2 2\theta_{12}}{\Delta^2_{21}}\sin\Phi_0\sum^{n-1}_{i=0}\Delta V_i \sin\Phi_i,
\end{equation}
where $n$ is the number of layers for the Earth mass density, $\Delta V_i=V_{i+1}-V_i$ is the potential difference between adjacent layers of matter, and $\Phi_i$ is the phase acquired along the trajectories. We take    $\sin^2\theta_{12} = 0.308$ and $\Delta^2_{21} = 7.54 \times 10^{-5}~\rm eV^{2}$, which are best fit values of neutrino mixing parameters from a recent global fitting \cite{Capozzi:2013csa}. For the rest of flavors, the condition of flux conservation gives
\begin{equation}
4F_x = F^0_e + F^0_{\bar{e}} + 4F^0_x - F_e - F_{\bar{e}} = F^0_e + \bar{P}_{2e} F^0_{\bar{e}} + (3-\bar{P}_{2e}) F^0_x,
\end{equation}
and
\begin{equation}
4F_x = F^0_e + F^0_{\bar{e}} + 4F^0_x - F_e - F_{\bar{e}} = (1-P_{2e}) F^0_e + F^0_{\bar{e}} + (2+P_{2e}) F^0_x,
\end{equation}
for the normal and inverted hierarchies, respectively.


\section{Spectrum for Inverse Beta Decay and Neutral Current Interaction}

From Eqs. (\ref{eNH}) to (\ref{ebarIH}), it is shown that, in the normal hierarchy, $\nu_e$ completely comes from $\nu_x^{0}$ from the source while $\bar{\nu}_e$ comes from both $\bar{\nu}_e^{0}$ and $\bar{\nu}_x^{0}$. On the other hand, in the inverted hierarchy, $\nu_e$ comes from both $\nu_e^{0}$ and $\nu_x^{0}$ while $\bar{\nu}_e$ completely comes from $\bar{\nu}_x^{0}$. It is also shown that Earth matter effects occur on $\bar{\nu}_e$ in the normal hierarchy and $\nu_e$ in the inverted hierarchy.

In this section, we discuss interactions of SN neutrinos in liquid scintillation detectors. In scintillation detectors, inverse beta decays (IBD) are the most dominant interactions. The yield of $\nu p$ elastic scatterings is also comparable to that of IBD due to the large number of free protons \cite{Dasgupta:2011wg}. The interaction spectra  are given as
\begin{eqnarray}
&&\left(\frac{dN}{dE_\nu}\right)_{\rm IBD}=N_p\cdot\frac{dF_{\bar{e}}}{dE_\nu}\cdot\sigma_{\rm IBD}(E_\nu), \label{IBDspec}  \\
&&\left(\frac{dN}{dE_\nu}\right)_{\rm NC}=N_p\cdot\int_0^{T_{\rm max}}\frac{dF_{\rm tot}}{dE_\nu}\frac{d\sigma_{\nu p}(E_\nu)}{dT}dT. \label{NCspec}
\end{eqnarray}
Here $N_p$ is the number of the target protons in the detector and $F_{\rm tot} \equiv F_e+F_{\bar{e}}+4F_x$  is the total fluence of the SN neutrinos. NC denotes $\nu p$ elastic scatterings for they are neutral current interactions.

We note that the interaction spectra, Eq. (\ref{IBDspec}) and (\ref{NCspec}), account for the number of IBD and NC interactions occurring inside the detector per energy of incident SN neutrinos. They are not the observed event spectra but can be constructed from the observed spectra.   

The observed event spectrum, $dN/dE_{e^+}$, for IBD is obtained in scintillation detectors by measuring the positron energy deposit. To a good approximation, one has $E_{e^+}=E_\nu - 1.3~{\rm MeV}$ for $E_\nu < 300~{\rm MeV}$ as indicated in \cite{Strumia:2003zx}. In the energy regime of SN neutrinos, this allows the event spectrum $dN/dE_{e^+}$ to be directly converted to the interaction spectrum $(dN/dE_\nu)_{\rm IBD}$ with cross section $\sigma_{\rm IBD}(E_\nu)$ taken from \cite{Strumia:2003zx}.

SN neutrinos interact with free protons in the detector through neutral current elastic scattering, producing protons with recoil kinetic energies $T$. To produce a proton recoil energy $T$ requires a minimum neutrino energy $E_{\nu,{\rm min}}=\sqrt{m_p T/2}$, with $m_p$ the proton mass. In other words, a neutrino of energy $E_\nu$ can produce a proton recoil energy between $0$ and $T_{\rm max}=2E^2_\nu/m_p$. These protons are slow hence they are detected with quenched energies $T'<T$. The proton recoil energy $T$ is mapped to an electron-equivalent quenched energy $T^{\prime}$ through the quenching function
\begin{equation}
T^{\prime}(T)=\int_0^T\frac{dT}{1+k_B<dT/dx>},
\end{equation}
where $k_B$ is Birks constant \cite{Birks:1951}. The observed event spectrum for NC interactions is actually the effective proton spectrum 
\begin{equation}
\frac{dN}{dT'}=\frac{N_p}{dT'/dT}\int_{E_{\nu,{\rm min}}}^\infty dE_\nu\frac{dF_{\rm tot}}{dE_\nu}\frac{d\sigma}{dT}(E_\nu).
\end{equation}
A measured $T'$ corresponds to a unique $T$ using the known quenching function. The energy $T$ is then related to $E_{\nu,{\rm min}}$ via $E_{\nu,{\rm min}}=\sqrt{m_p T/2}$ mentioned before. Once the quenched spectrum $dN/dT'$ is measured, one can extract the neutrino fluence $dF/dE$ with the inversion process described in \cite{Dasgupta:2011wg}. The spectrum for neutral current interactions, $(dN/dE_\nu)_{\rm NC}$, is then given by Eq. (\ref{NCspec}).

Besides IBD and NC signals, there are still other interactions between SN neutrinos and scintillation materials. Neutrinos of all flavors also interact with electrons via $\nu + e^- \rightarrow\nu + e^-$. The recoil kinetic energies of the scattered electrons range from $0$ to $2E^2_\nu/(2E_\nu+m_e)$. Unlike the proton recoil energy, the electron recoil energy is not quenched. These $\nu e$ elastic scatterings can be detected directly by measuring the recoiled signals \cite{Laha:2013hva,Laha:2014yua}. Meanwhile, neutrinos also interact with carbon nuclei. By neutral current interactions, neutrinos can excite ground state $^{12}{\rm C}$. The excited carbon nucleus soon jumps back to the ground state, spontaneously emitting a $15.11~{\rm MeV}$ photon. By charged current interactions with $^{12}{\rm C}$, electron neutrinos produce $^{12}{\rm N}_{\rm g.s.}$:
\begin{equation}
\nu_e + {^{12}{\rm C}}\rightarrow {^{12}{\rm N}}_{\rm g.s.} + e^-,
\end{equation} 
and electron antineutrinos produce $^{12}{\rm B}_{\rm g.s.}$:
\begin{equation}
\bar{\nu}_e + {^{12}{\rm C}}\rightarrow {^{12}{\rm B}}_{\rm g.s.} + e^+,
\end{equation} 
where g.s. denotes the ground state. Both $^{12}{\rm N}_{\rm g.s.}$ and $^{12}{\rm B}_{\rm g.s.}$ decay back to $^{12}{\rm C}$ within tens of milliseconds and emit a positron with a maximum kinetic energy of $\approx16.8~{\rm MeV}$ and an electron with a maximum kinetic energy of $\approx12.9~{\rm MeV}$, respectively. Though it is difficult to distinguish between these two kinds of charged current events, both channels can be detected by the space and time coincidence of the scattering and its following decay.

In addition to these neutrino interactions with carbon nuclei, which can be distinguished from recoil signals by identifying their unique characteristics, neutrinos can also produce protons by interacting with carbon nuclei by so-called proton knockouts \cite{Lujan-Peschard:2014lta},
\begin{eqnarray}
\nu (\bar{\nu})+ {^{12}{\rm C}} & \rightarrow & {^{11}{\rm B}} + p + \nu (\bar{\nu}),  \\
\nu  + {^{12}{\rm C}}  &  \rightarrow &  {^{11}{\rm C}} + e^- + p.
\end{eqnarray}
We note that the $\nu e$ scattering signals can mix with the proton recoil signals. The range of the electron recoil energy from $0$ to $\sim30~{\rm MeV}$ is broader than the range of the proton recoil energy from $0$ to $\sim5~{\rm MeV}$ while their events are comparable. Moreover, the proton recoil energy inside the scintillator is further quenched to $\lesssim2~{\rm MeV}$. The $\nu e$ scattering, together with proton knockout interactions only make up a small fraction of $<10\%$ of the signals within the energy range of quenched proton recoils \cite{Lujan-Peschard:2014lta}. For simplicity, we neglect it in this paper.

We point out that not all signals within the energy range of proton recoils are taken into account. Since the scintillator is made of hydrocarbon, a natural isotope of the carbon, $^{14}{\rm C}$, decays into $^{14}{\rm N}$, emitting electrons below $0.2$ MeV with a high rate. Below this energy, the signal is flooded by the above background electrons. Therefore, a threshold of $T^{\prime}=0.2~{\rm MeV}$ is set for recording the signal. The threshold of $T^{\prime}$ is converted to the threshold of proton recoil energy $T$ and the corresponding $E_{\nu,{\rm min}}$. Signals are recorded only for neutrinos with energies above this $E_{\nu,{\rm min}}$. Therefore, only the higher energy part of the neutrino fluence is reconstructed as demonstrated in \cite{Dasgupta:2011wg}. 
\label{NCreconstruct}

\section{Resolving Neutrino Mass Hierarchy}

\begin{figure}[htbp]
	\begin{center}
	\includegraphics[width=8cm]{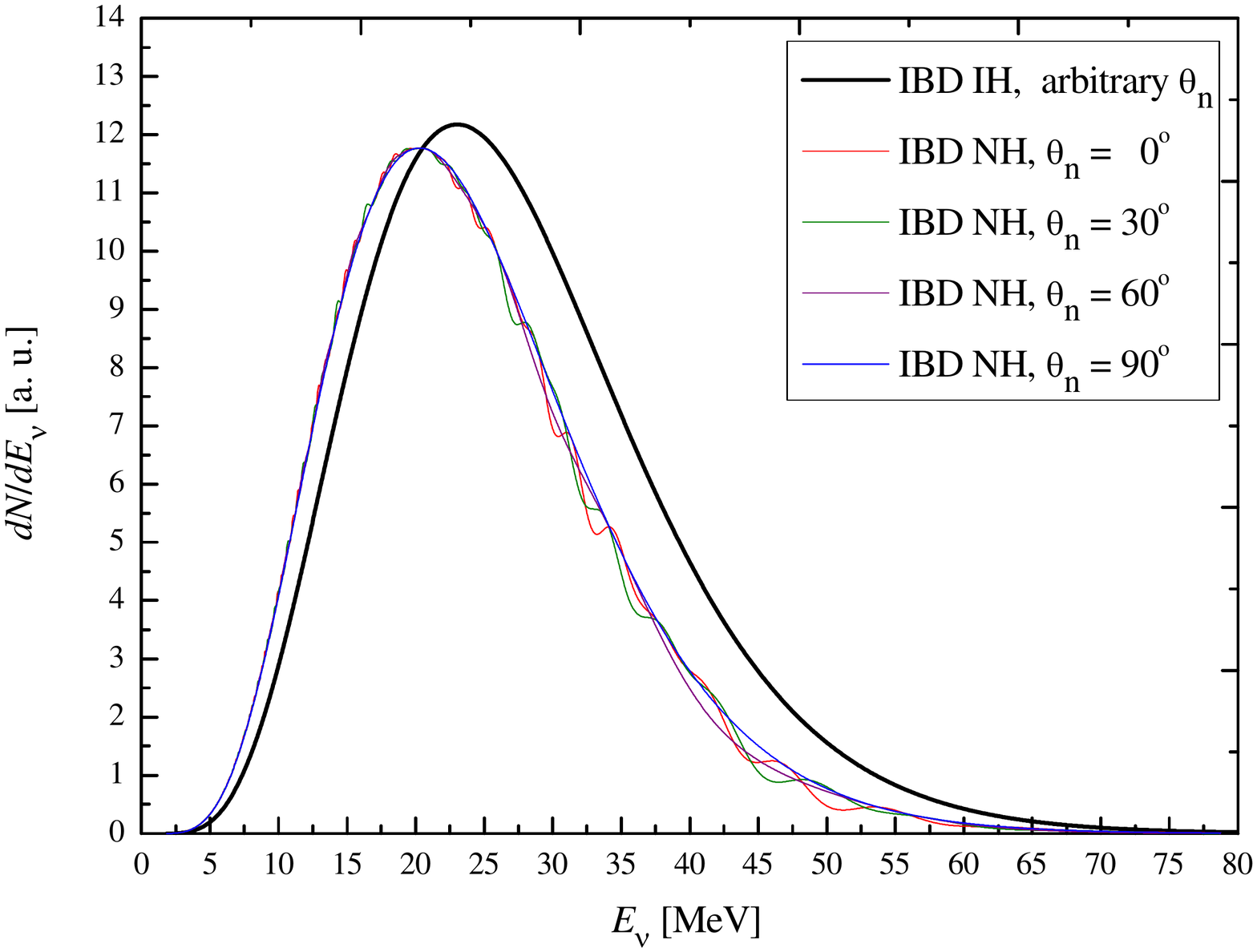}
	\caption{IBD interaction spectra for normal and inverted mass hierarchies, respectively with different nadir angles. A crossover occurs at $E_\nu\sim20~{\rm MeV}$, above which the spectrum for the inverted hierarchy becomes larger than all the spectra for the normal hierarchy despite of the Earth matter effect.}
	\label{fig:IBD}
	\end{center}
\end{figure}

Since the $\nu p$ elastic scattering cross section is identical for all flavors and accounts for the total neutrino fluence, no difference appears in the NC spectrum with respect to the neutrino mass hierarchy. Hence, detecting only NC interactions cannot distinguish neutrino mass hierarchies. However, the dense matter inside the SN and the Earth matter effect would shape the spectrum of the $\bar{\nu}_e$ fluence in a way depending on the neutrino mass hierarchy. As shown in Eqs. (\ref{ebarNH}) and (\ref{ebarIH}), the $\bar{\nu}_e$ fluence is modified by the matter effect as neutrinos propagate outwards through the dense medium inside the SN for both hierarchies. On the other hand, the Earth matter effect only affects the $\bar{\nu}_e$ fluence in the normal mass hierarchy. Consequently, the IBD spectra are sensitive to neutrino mass hierarchy as shown in Fig. \ref{fig:IBD}. First, due to Earth matter effects, fluctuations occur in the IBD spectrum for the normal hierarchy while the IBD spectrum remains smooth for the inverted mass hierarchy. Second, a crossover appears at approximately $\sim20{\rm MeV}$, such that the IBD spectrum for the normal hierarchy is larger than that for the inverted hierarchy and it is opposite below this energy. In other words, for energies larger than $\sim20{\rm MeV}$, there are more IBD events in the inverted mass hierarchy than those in the normal hierarchy.

\begin{figure}[htbp]
	\begin{center}
	\includegraphics[width=8cm]{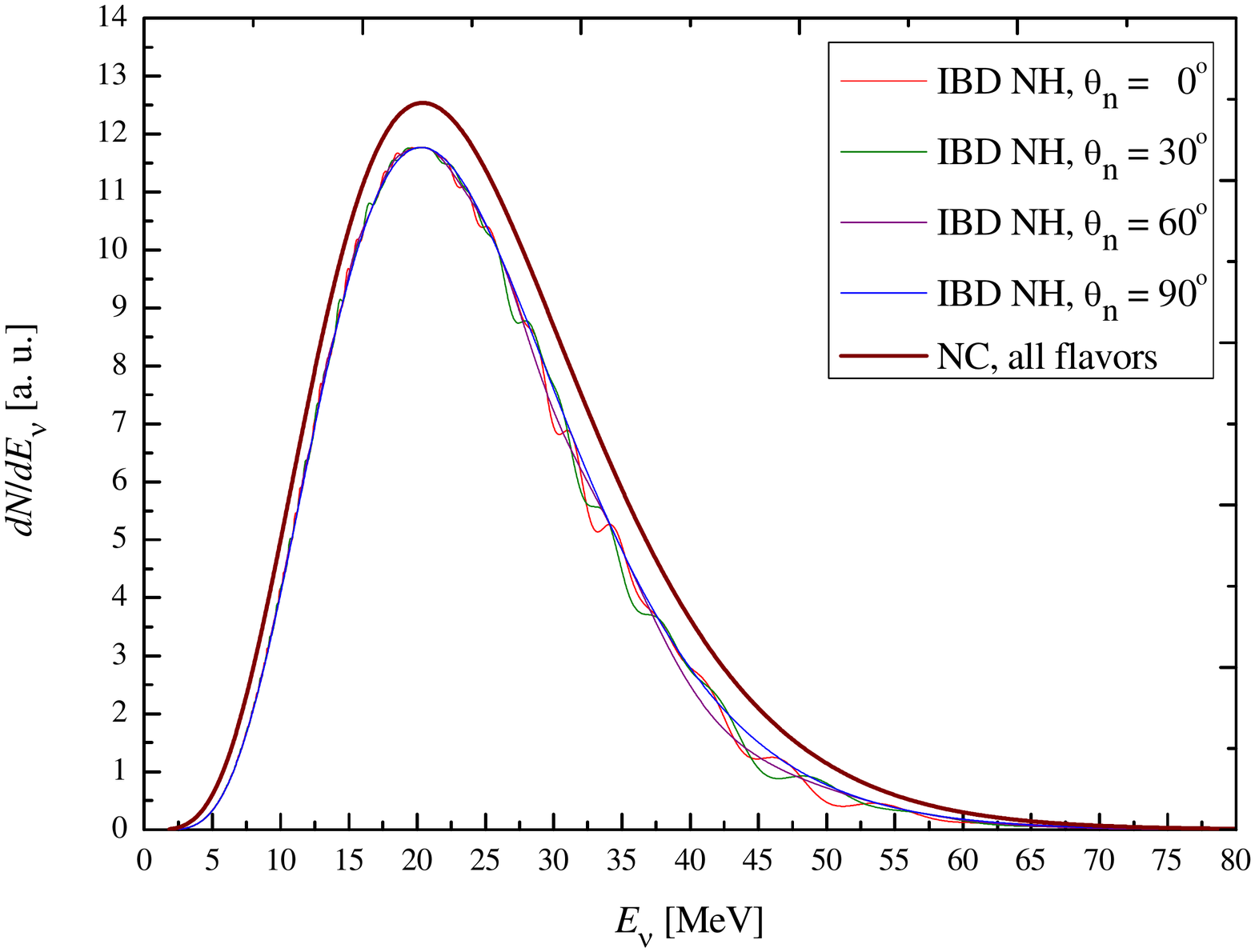}
	\includegraphics[width=8cm]{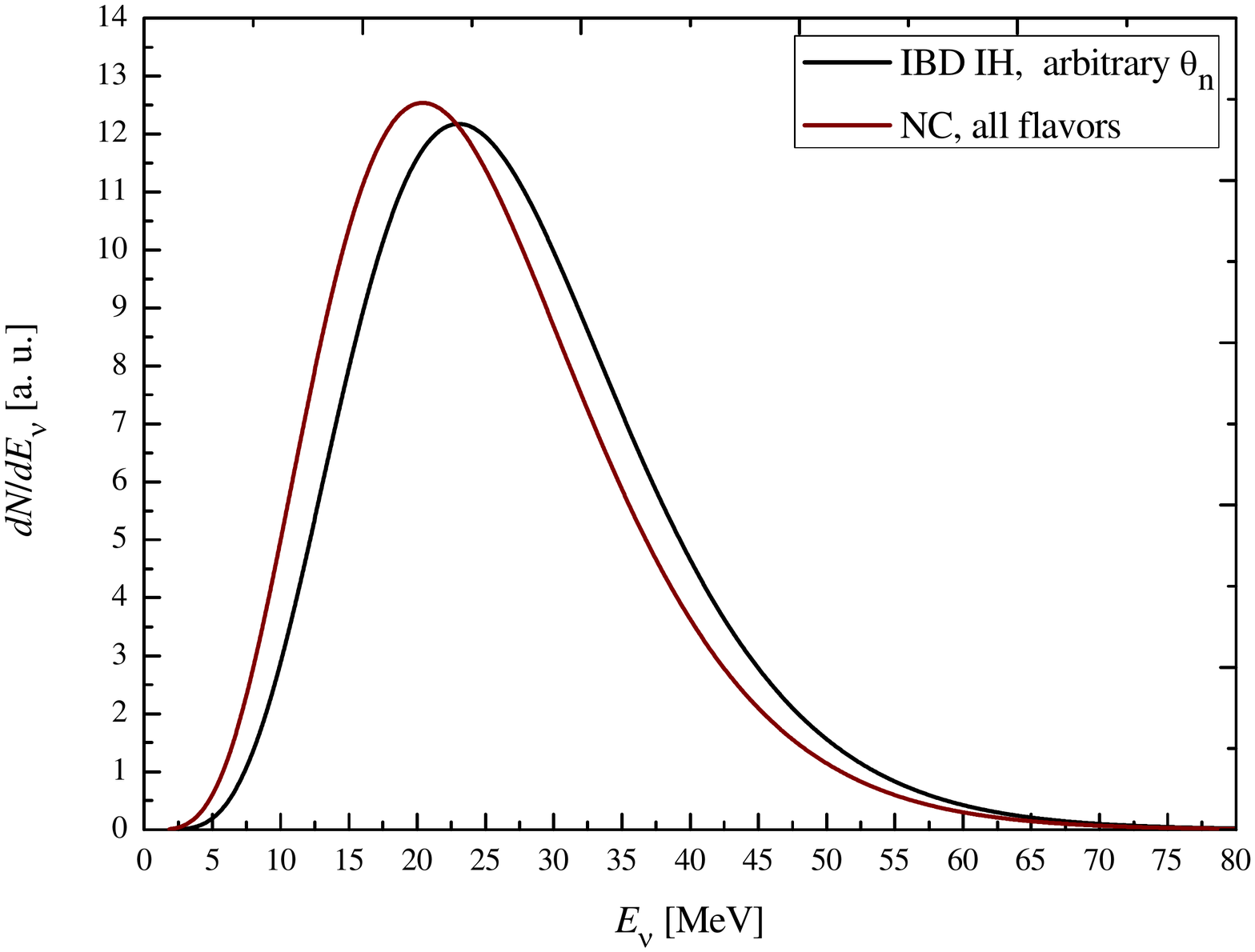}
	\caption{The comparison between NC and IBD interaction spectra. In the left panel, NC interaction spectrum is higher than the IBD one in the normal hierarchy over the entire energy range regardless the nadir angle of the incident neutrinos. In the right panel, NC interaction spectrum becomes lower than the IBD one in the inverted hierarchy above the crossover energy around $25~{\rm MeV}$.}
	\label{fig:interaction}
	\end{center}
\end{figure}

Clearly, the above two features of the IBD spectrum suggest two different approaches for identifying the neutrino mass hierarchy. The first approach studies the energy distribution of IBD events. The second approach relies on probing the ripples in the IBD spectrum, arising from the Earth matter effect. In this work, we take the first approach by using NC interaction spectrum to weigh the IBD events.

\begin{figure}[htbp]
	\begin{center}
	\includegraphics[width=8cm]{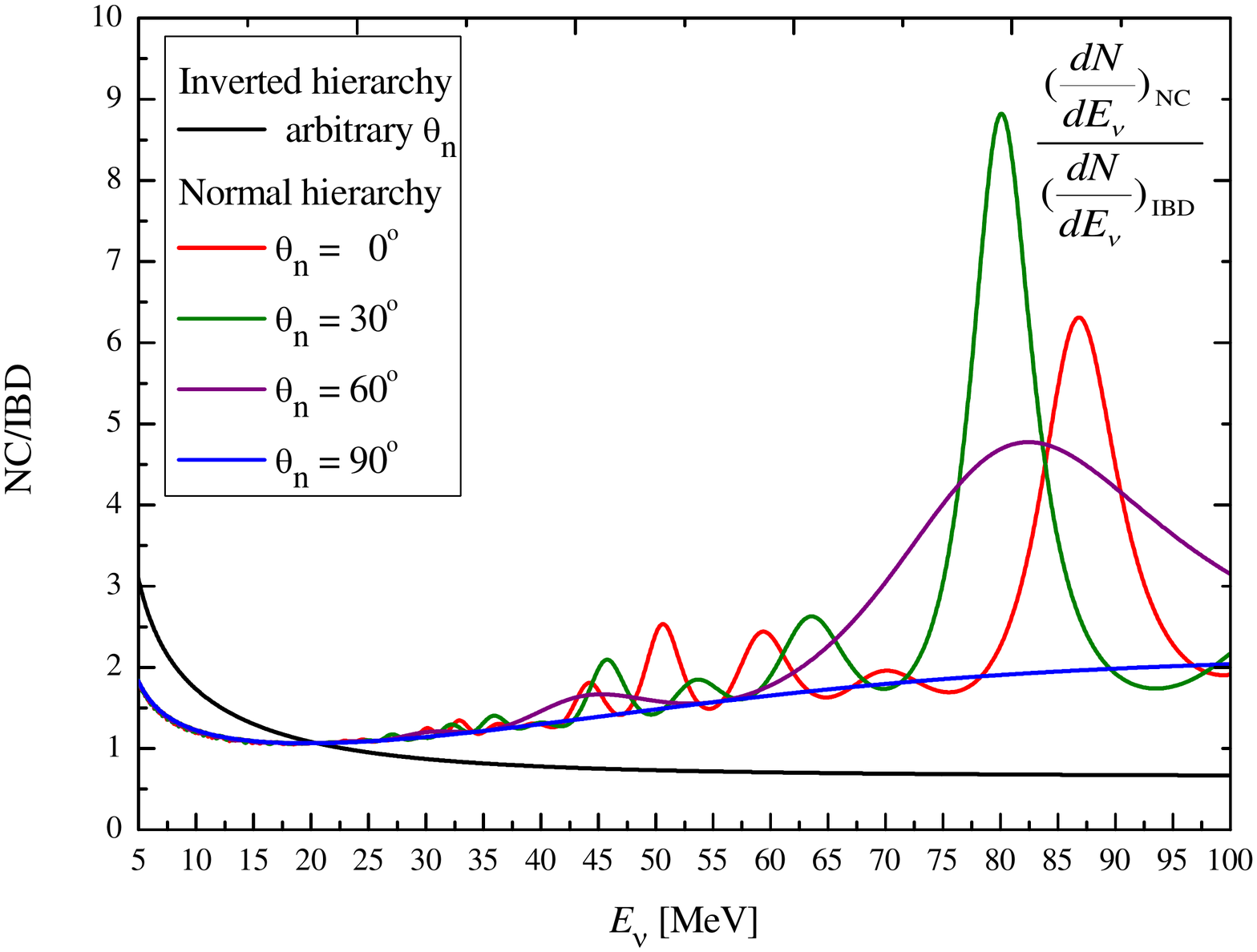}
	\includegraphics[width=8cm]{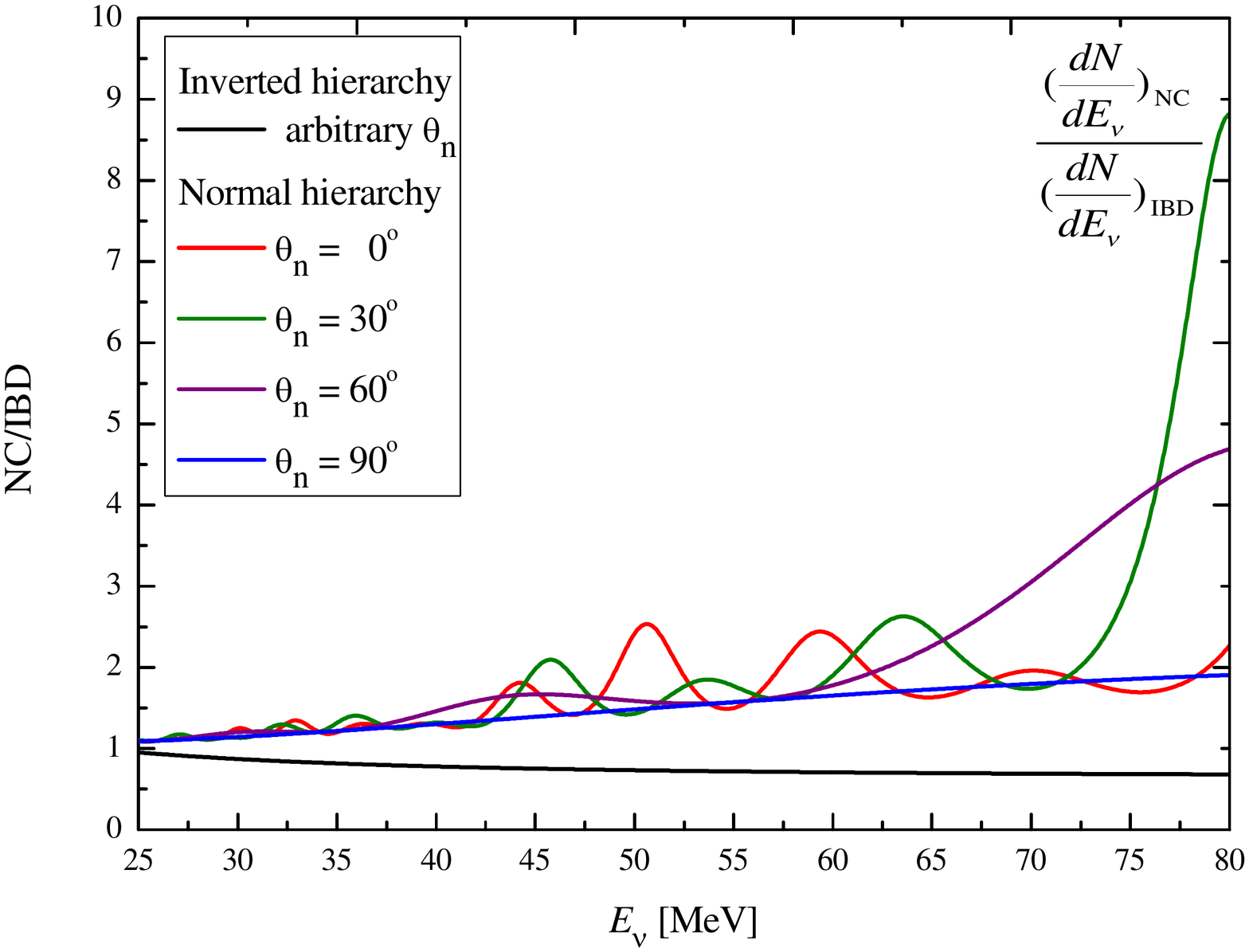}
	\caption{Ratio of NC interactions to IBD interactions. The left panel shows the NC to IBD ratio for $E_\nu$ between 5 MeV and 100 MeV while the right panel focuses on the energy range between 25 MeV and 80 MeV.}
	\label{fig:specratio}
	\end{center}
\end{figure}

On the left panel of Fig. \ref{fig:interaction}, one can see that the spectrum of NC is higher than that of IBD in the entire energy range for the normal hierarchy. On the right panel, it is seen that the spectrum of NC remains higher than that of IBD at low energies and becomes lower than that of IBD for $ E_\nu> 20~{\rm MeV}$ for the inverted hierarchy. This observation inspires us to define the number ratio of NC interactions to IBD interactions. As shown in Fig. \ref{fig:specratio}, the spectrum ratio of NC to IBD is always larger than one with minimum occurring at the energy of $\sim 20~{\rm MeV}$ for the normal hierarchy and is monotonic decreasing with energy and equals to one at the energy of $\sim 20~{\rm MeV}$ for the inverted hierarchy. Here we do not need to use the Earth matter effect for identifying the neutrino mass hierarchy. Instead, we probe the ratio of total NC interactions to total IBD interactions for neutrino energies higher than $20~{\rm MeV}$.

Once the interaction spectra of IBD and NC interactions, denoted as $(dN/dE_\nu)_{\rm IBD}$ and $(dN/dE_\nu)_{\rm NC}$, respectively, are obtained from measurements, one can calculate total interactions of both channels for $E_\nu>E_{\nu,{\rm s}}$, with $E_{\nu,{\rm s}}$ the selected energy cut, which must be larger than the crossover energy on the left panel of Fig. \ref{fig:specratio}. We define $R$ to be the ratio of the total interactions of NC to those of IBD for $E_\nu>E_{\nu,{\rm s}}$ i.e.,
\begin{equation}
R=\frac{\int_{E_{\nu,{\rm s}}}^\infty \left(\frac{dN}{dE}\right)_{\rm NC}dE}{\int_{E_{\nu,{\rm s}}}^\infty \left(\frac{dN}{dE}\right)_{\rm IBD}dE}.
\end{equation}
We study $R$ for  scintillation detectors such as  Borexino, KamLAND, SNO$+$, JUNO, and LENA.

\begin{table}[htbp]
\begin{center}
\begin{tabular}{lccccccc}\hline\hline
                  & Mass  &            $N_{\rm p}$ & $k_B$      &  Expected Events           &   $T$  & $E_{\nu,{\rm min}}$   \\ 
                  & [kton]   &           $[10^{31}]$  & [cm/MeV] & $T^{\prime}>0.2~{\rm MeV}$ & [MeV] &  [MeV]   \\ \hline
Borexino    & 0.278   &         1.7       &     0.010     &         27        &  1.00  &  21.67     \\ 
KamLAND & 0.697  &            5.9      &     0.010     &         66         &  1.33  &  25.01      \\
SNO$+$    & 0.800  &            5.9      &     0.0073   &        111        &   0.86  &  20.11          \\
JUNO        &  20 &           144      &     0.00759  &       2490      &   0.93 &   20.84       \\
LENA         &  44 &            325     &     0.010      &        5060      &  1.02  &  21.82          \\ \hline

\end{tabular}
\end{center}
\caption{NC events expected for various scintillation detectors. Chemical compositions and masses of scintillation materials, corresponding Birks constants ($k_{\rm B}$), numbers of free protons ($N_{\rm p}$), thresholds of proton recoil ($T$), thresholds of minimum neutrino energy ($E_{\nu,{\rm min}}$),  and expected numbers of NC events for the detectors are listed here.} 
\label{NCevent}
\end{table}

Since NC events are well measured for energies higher than the threshold $E_\nu>E_{\nu,{\rm min}}$, we shall only take higher energy part of NC and IBD interaction spectra for calculating the ratio $R$. Hence the energy cut $E_{\nu,{\rm s}}$ should be greater than the threshold energy $E_{\nu,{\rm min}}$. We note that the threshold energy $E_{\nu,{\rm min}}$ depends on the scintillation material since the minimum of the proton recoil energy $T$ corresponding to the threshold $T^{\prime}=0.2~{\rm MeV}$ depends on the Birks constant $k_{\rm B}$. In Table \ref{NCevent}, we list the parameters of Borexino, KamLAND, SNO$+$, JUNO, and LENA detectors. We propose to measure $R$ with these detectors.  For different detectors, the threshold energy $E_{\nu,{\rm min}}$ varies from 20 MeV to 25 MeV. Therefore, we take $E_{\nu,{\rm s}}=25~{\rm MeV}$. Such an energy is also higher than the crossover energy indicated on the left panel of Fig. \ref{fig:specratio}.

\begin{table}[htbp]
\begin{center}
\begin{tabular}{lrrrrcccccc}\hline\hline
                  & \multicolumn{1}{c}{NC}  & \multicolumn{3}{c}{IBD} & \multicolumn{3}{c}{R} & \multicolumn{3}{c}{$\sigma_{\rm R}[10^{-2}]$} \\ \cline{3-11}
                  &        & IH & NH$~0^\circ$ & NH$~90^\circ$ & IH & NH$~0^\circ$ & NH$~90^\circ$   & IH & NH$~0^\circ$ & NH$~90^\circ$ \\ \hline
Borexino    & 40   &   48   &  32  &  33  & 0.83  & 1.25  & 1.21  & 20.0 & 32.7 & 31.5    \\ 
KamLAND     & 141  &  169   & 112  & 116  & 0.83  & 1.26  & 1.22  & 12.1 & 19.5 & 18.7    \\
SNO$+$      & 141  &  167   & 111  & 115  & 0.84  & 1.27  & 1.23  & 10.3 & 17.1 & 16.3        \\
JUNO        & 3413 &  4092  & 2713 & 2821 & 0.83  & 1.26  & 1.21  & 2.12 & 3.49 & 3.33     \\
LENA        & 7677 &  9204  & 6104 & 6345 & 0.83  & 1.26  & 1.21  & 1.46 & 2.39 & 2.28     \\ \hline

\end{tabular}
\end{center}
\caption{Numbers of NC and IBD interactions in different scintillation detectors} 
\label{ratios}
\end{table}

The right panel of Fig. \ref{fig:specratio} shows that, for $E_{\nu}>25~{\rm MeV}$, $(dN/dE)_{\rm NC} < (dN/dE)_{\rm IBD}$ for the inverted hierarchy while $(dN/dE)_{\rm NC} > (dN/dE)_{\rm IBD}$
for the normal hierarchy regardless the Earth matter effect. Therefore, we obtain $R>1$ for the normal hierarchy despite of earth matter effects and $R<1$ for the inverted hierarchy. We calculate total NC and IBD interactions and their ratios assuming a representative supernova described in Sec. \ref{fluence}. The $R$ values for different scintillation detectors such as Borexino, KamLAND, SNO$+$, JUNO, and LENA are listed in Table \ref{ratios}. For the inverted hierarchy, $R\cong0.84$ while $R\cong1.21-1.27$ for the normal hierarchy. The variation arises from the earth matter effect which depends on the nadir angle of incoming SN neutrinos.

\begin{figure}[htbp]
	\begin{center}
	\includegraphics[width=8cm]{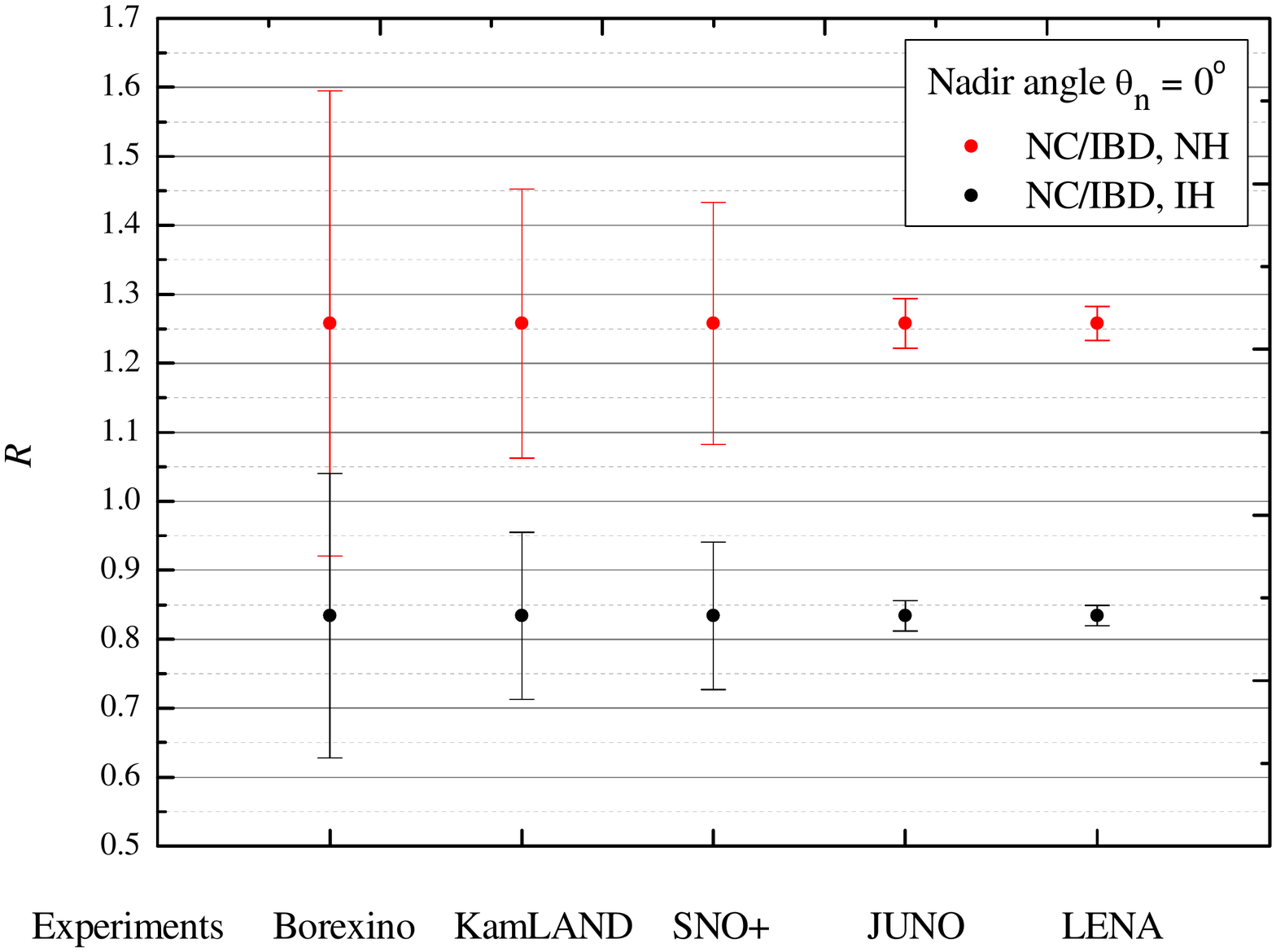}
	\includegraphics[width=8cm]{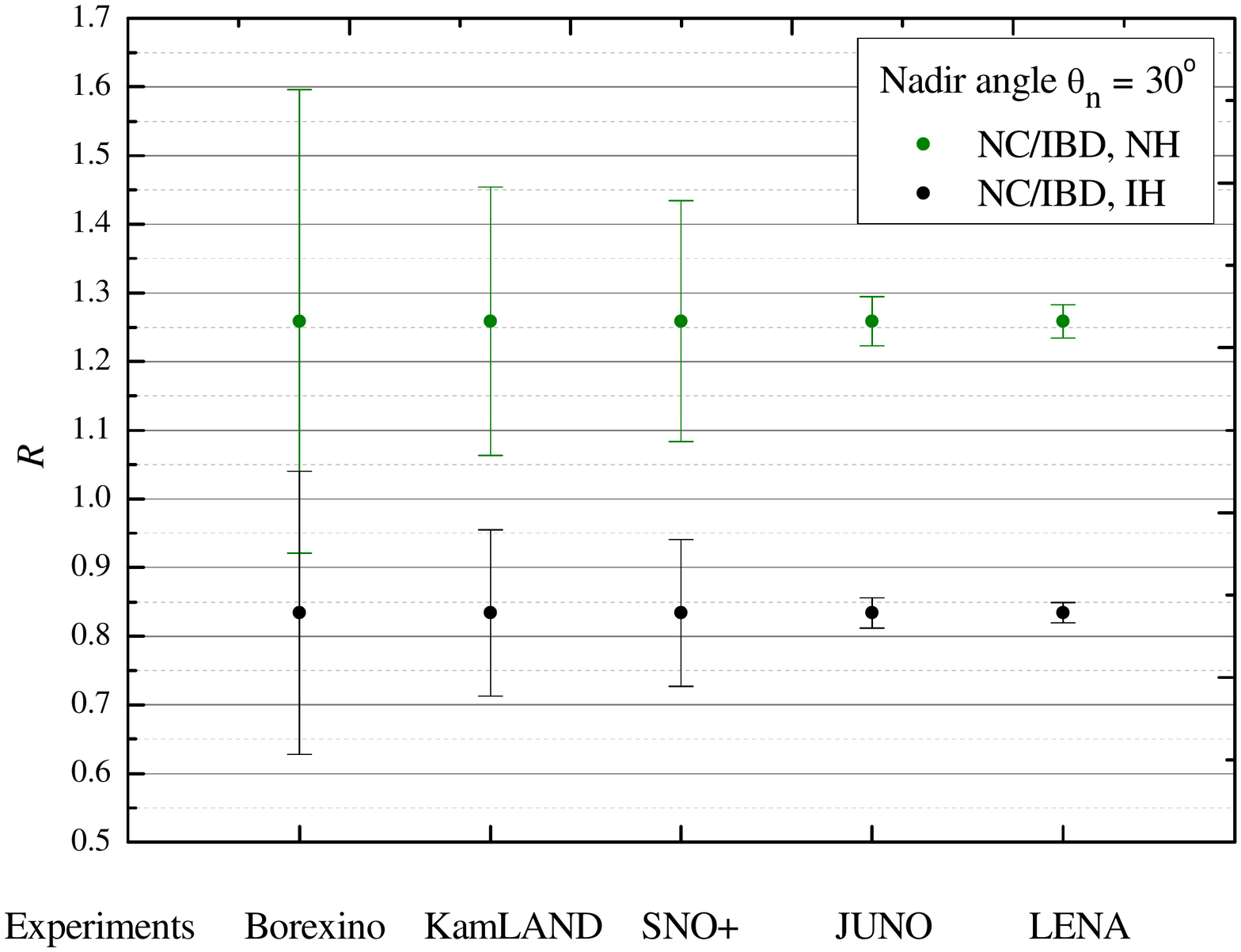}
	\includegraphics[width=8cm]{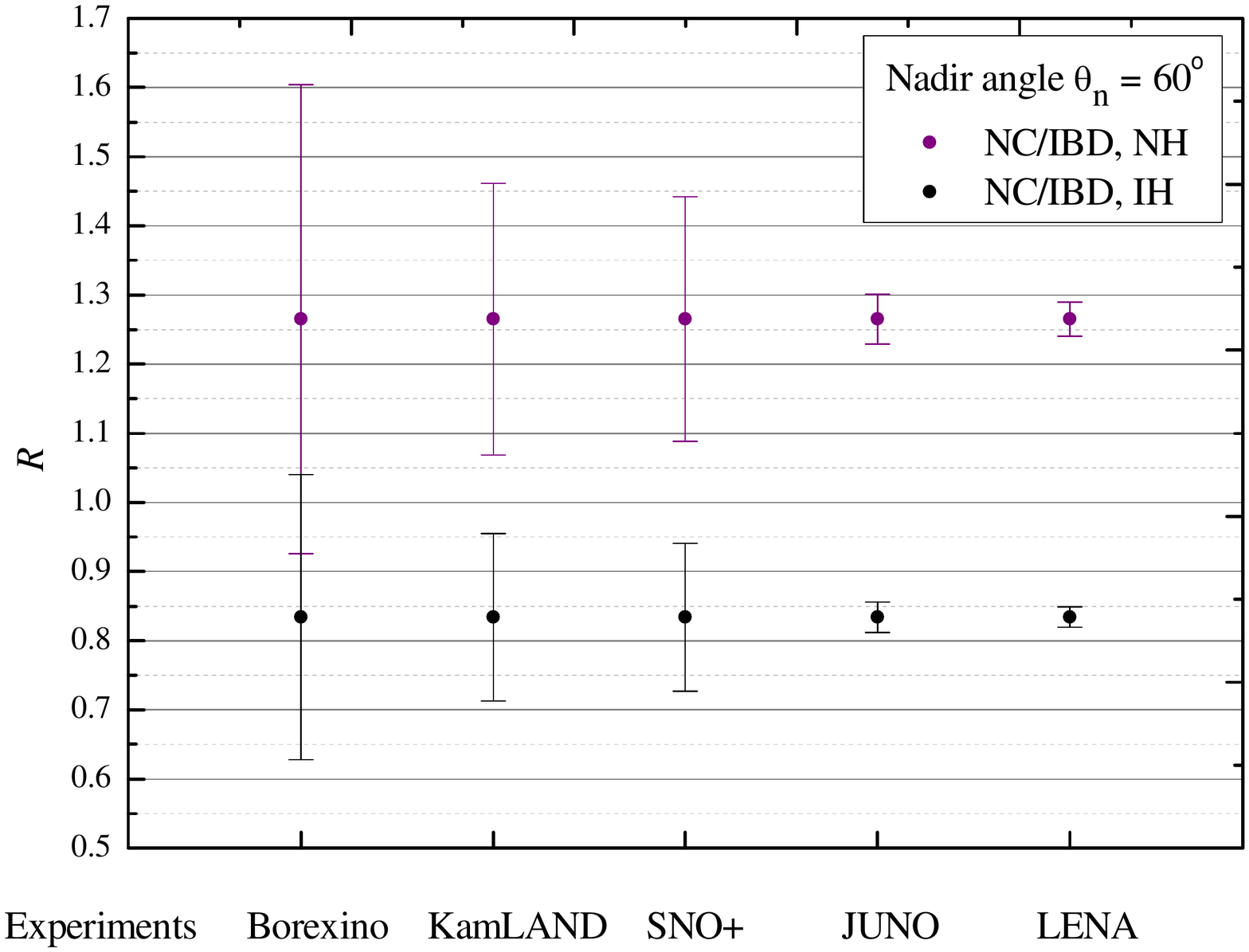}
	\includegraphics[width=8cm]{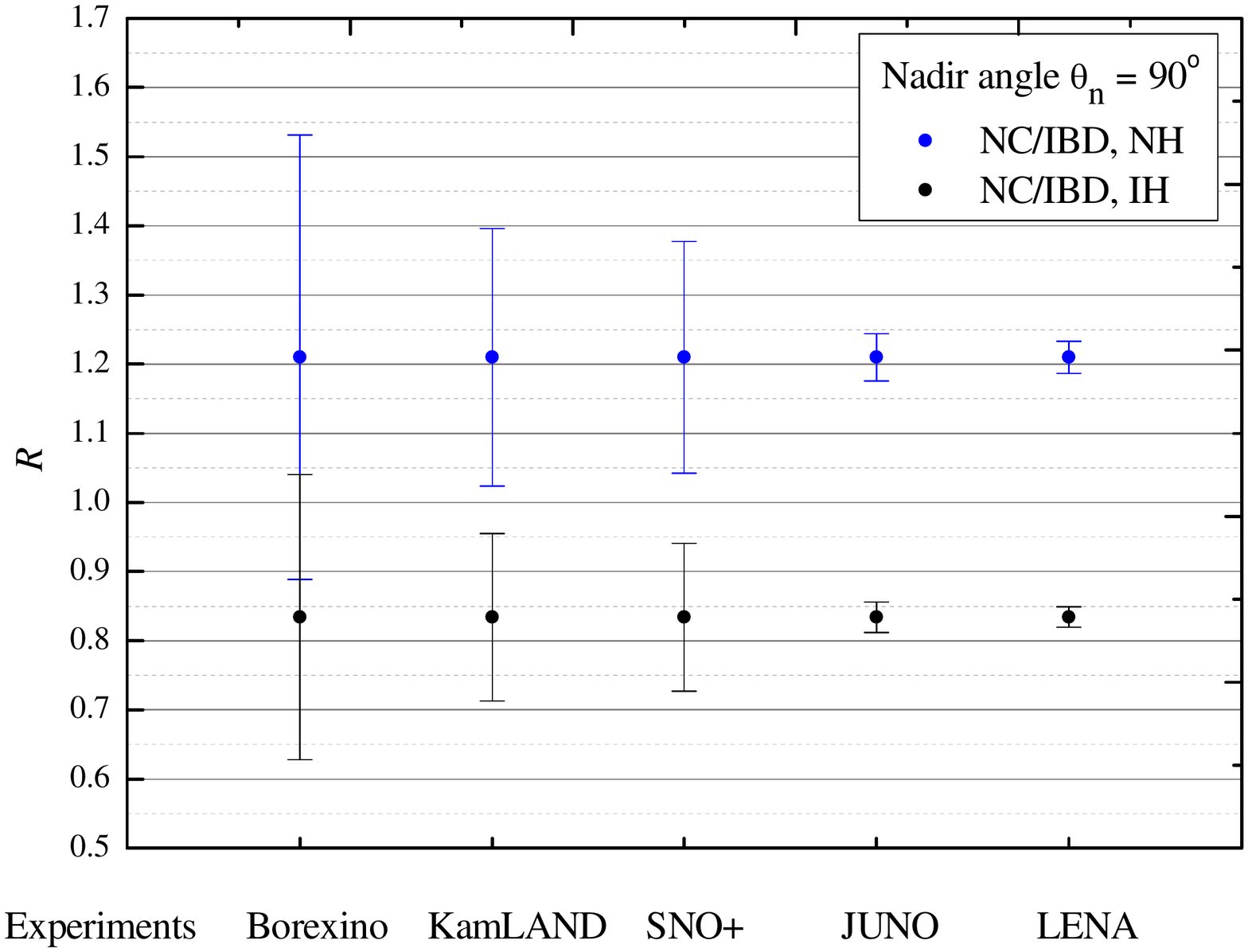}
	\caption{Ranges of $R$ for normal and inverted hierarchies at five scintillation detectors. }
	\label{fig:exptR}
	\end{center}
\end{figure}

We assume Poisson statistics for the expected number of events. The uncertainty $\sigma_R$ comes from the fluctuations of NC and IBD interactions. We note that the uncertainties of NC interactions are derived from the expected number of events listed in Table \ref{NCevent}. In addition to the statistical uncertainty, the energy resolution and quenching can also induce uncertainties in $R$. The energy resolution, $\Delta T^{\prime}/T^{\prime}$, are $6.9\%/\sqrt{T^{\prime}}$ for KamLAND and $\sim3\%/\sqrt{T^{\prime}}$ expected for JUNO. Uncertainties occur when recording signals around the threshold. For KamLAND, this increases $\sigma_R$ by about $7\%$. The quenching factor of KamLAND can vary in the range $k_B=0.0100\pm0.0002~{\rm cm/MeV}$. The effect of varying $k_B$ introduces less than $1\%$ effect on $\sigma_R$. The ranges of $R$ for different detectors are presented in Fig. \ref{fig:exptR}. The ranges of $R$ for normal and inverted hierarchies overlaps only for Borexino detector. For other detectors, the ranges are well-separated, indicating that the neutrino mass hierarchy can be resolved by these detectors, except Borexino, with neutrinos coming from a typical supernova.

\begin{figure}[htbp]
	\begin{center}
	\includegraphics[width=8cm]{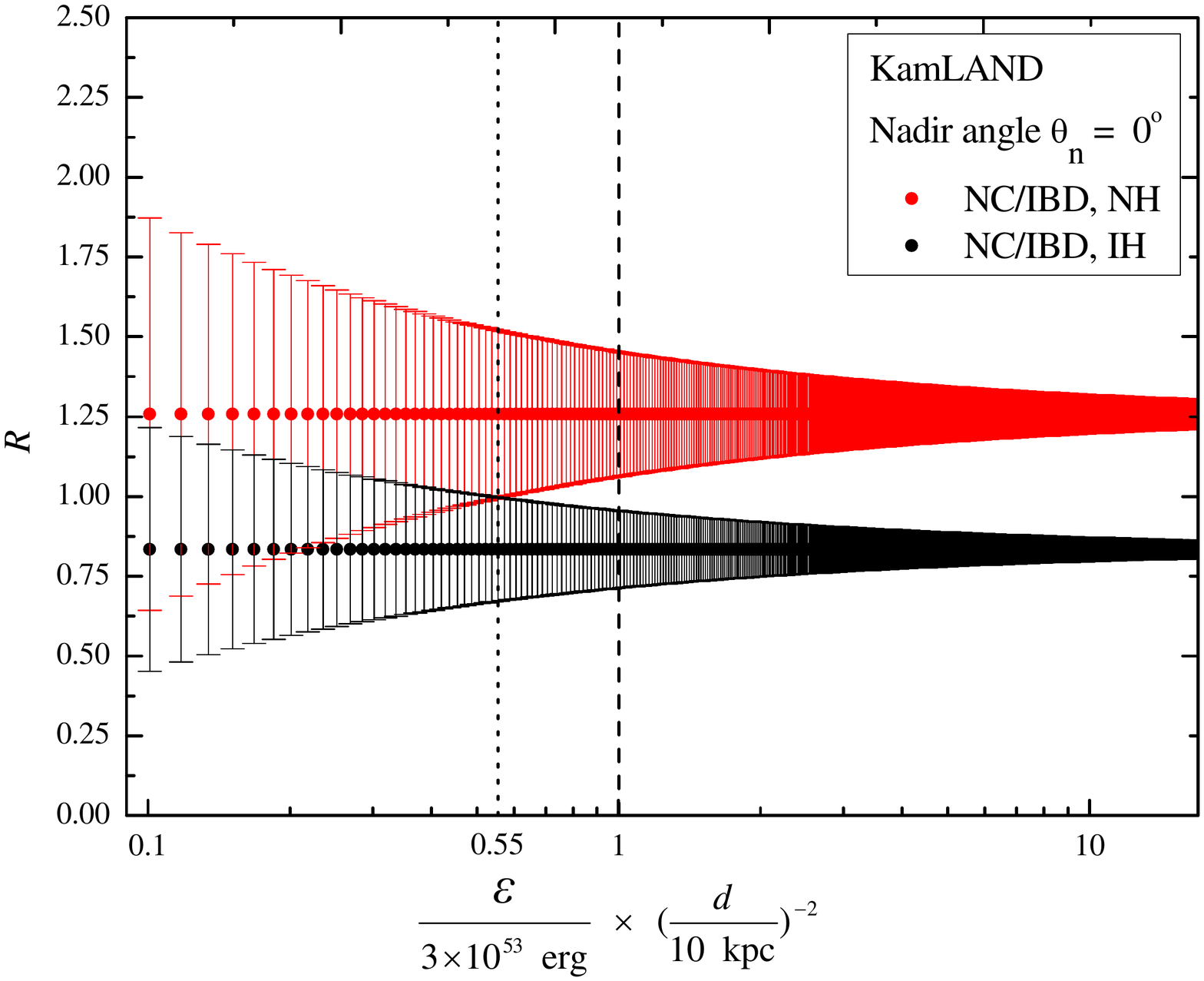}
    \includegraphics[width=8cm]{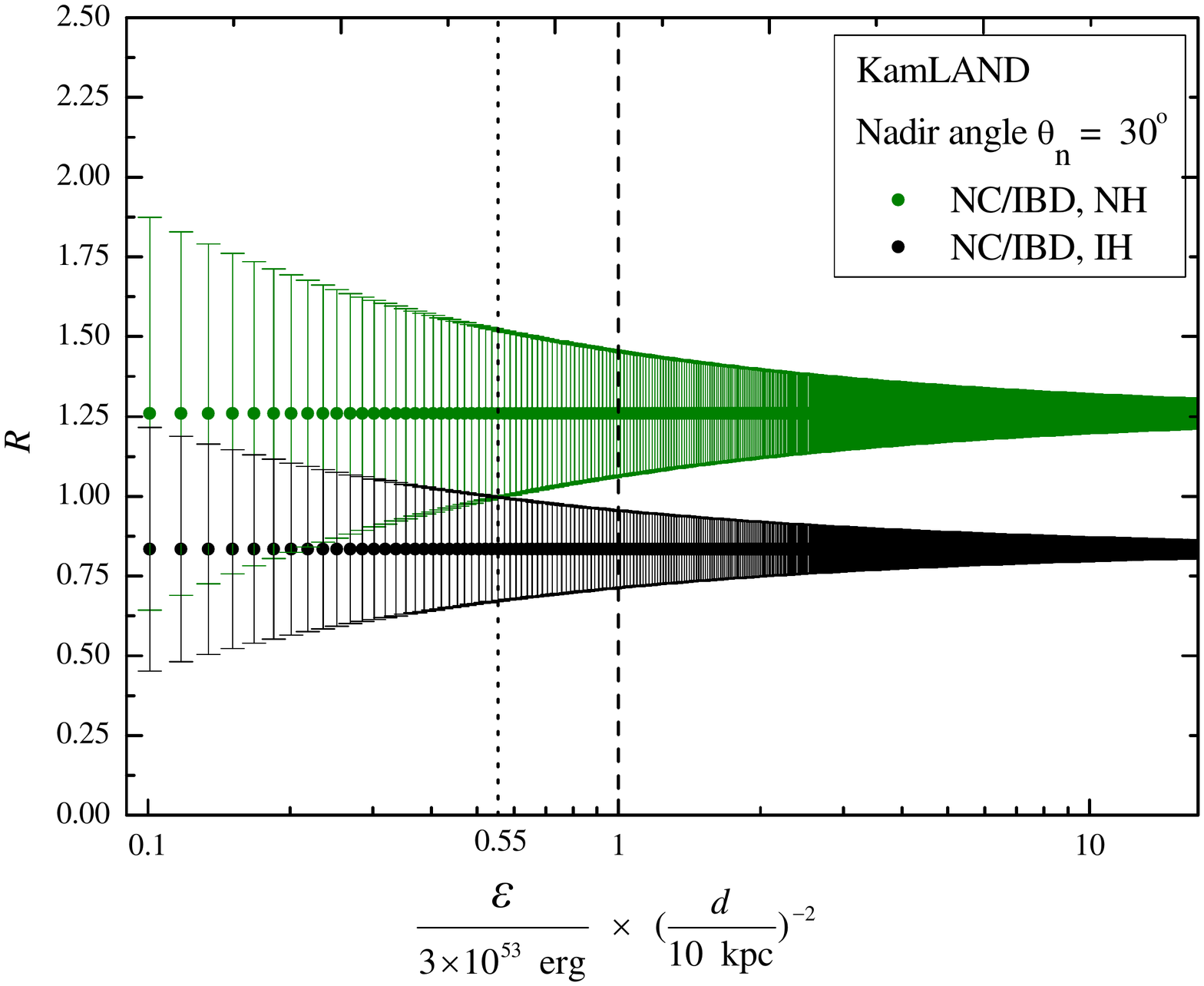}
    \includegraphics[width=8cm]{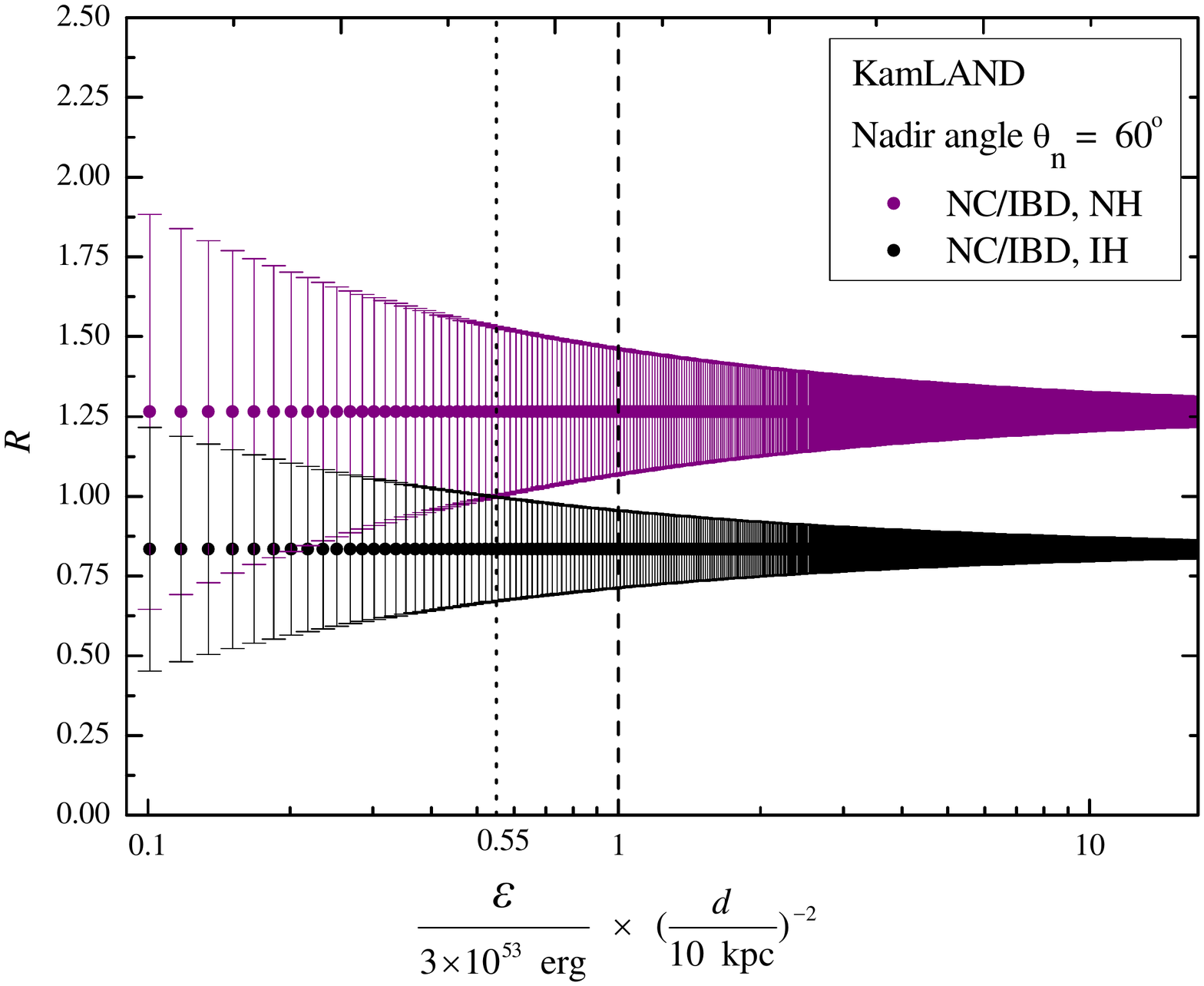}
	\includegraphics[width=8cm]{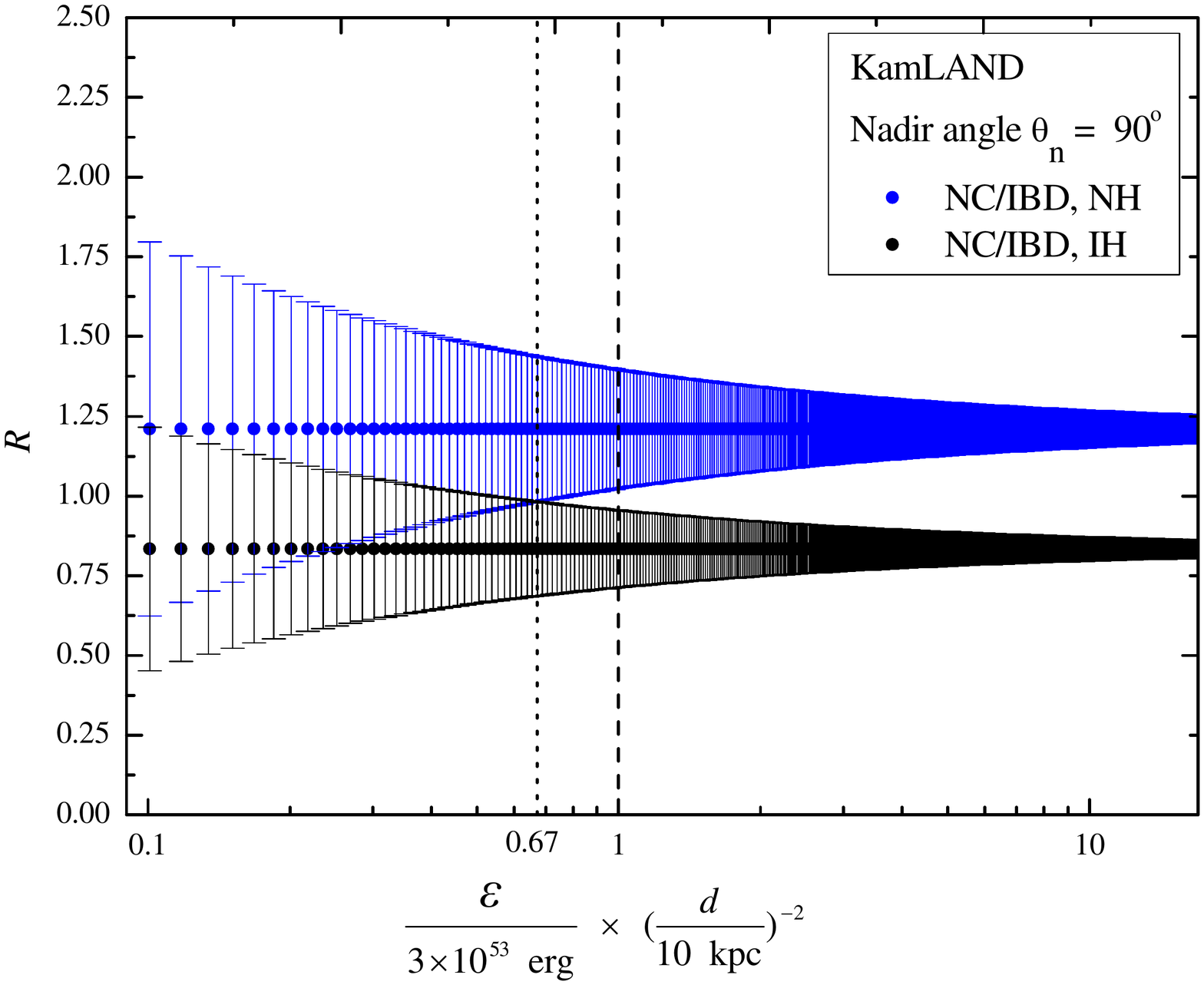}
	\caption{Ranges of $R$ for normal and inverted hierarchies expected at KamLAND detector}
	\label{fig:projectedR}
	\end{center}
\end{figure}

\begin{figure}[htbp]
	\begin{center}
	\includegraphics[width=8cm]{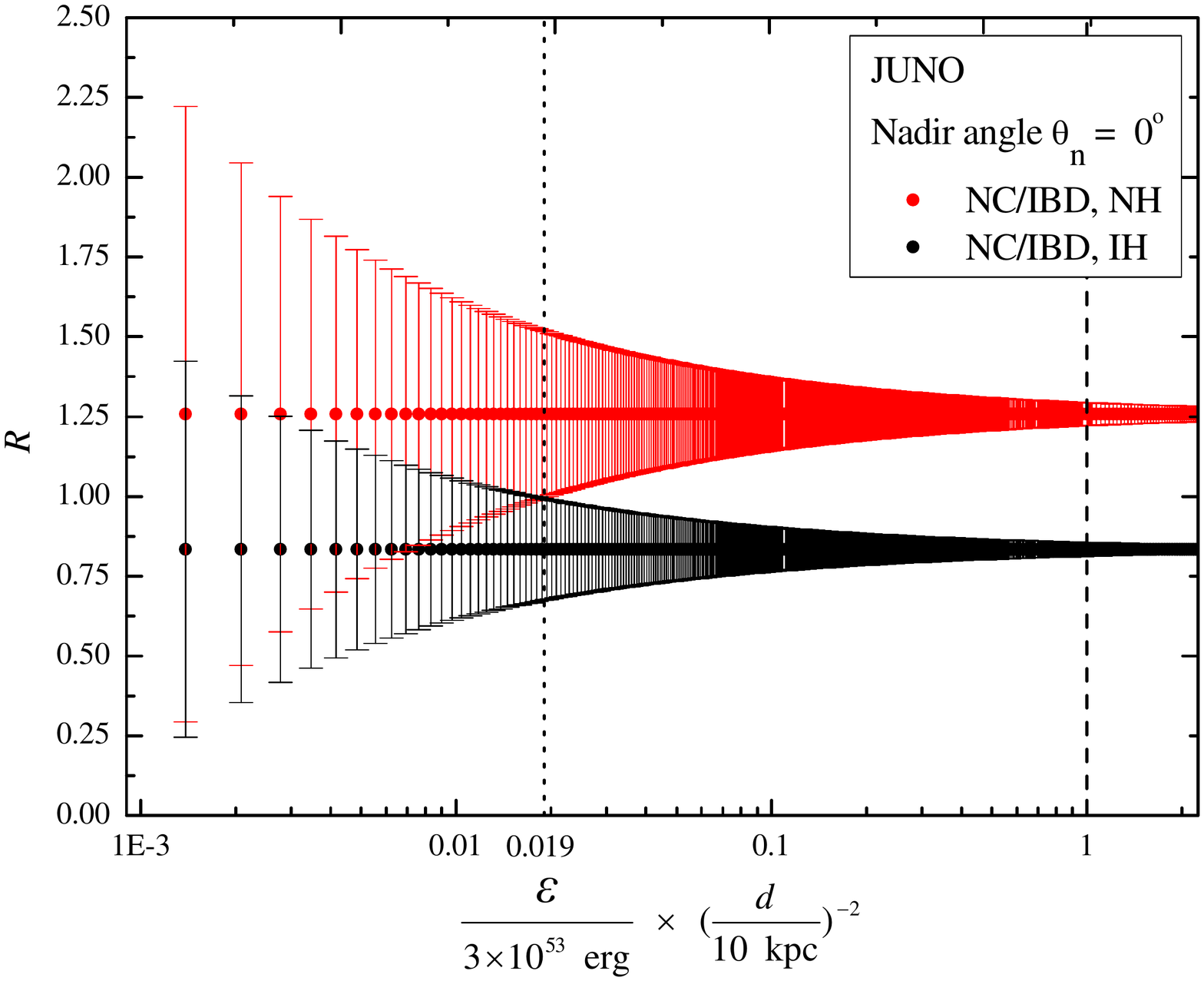}
	\includegraphics[width=8cm]{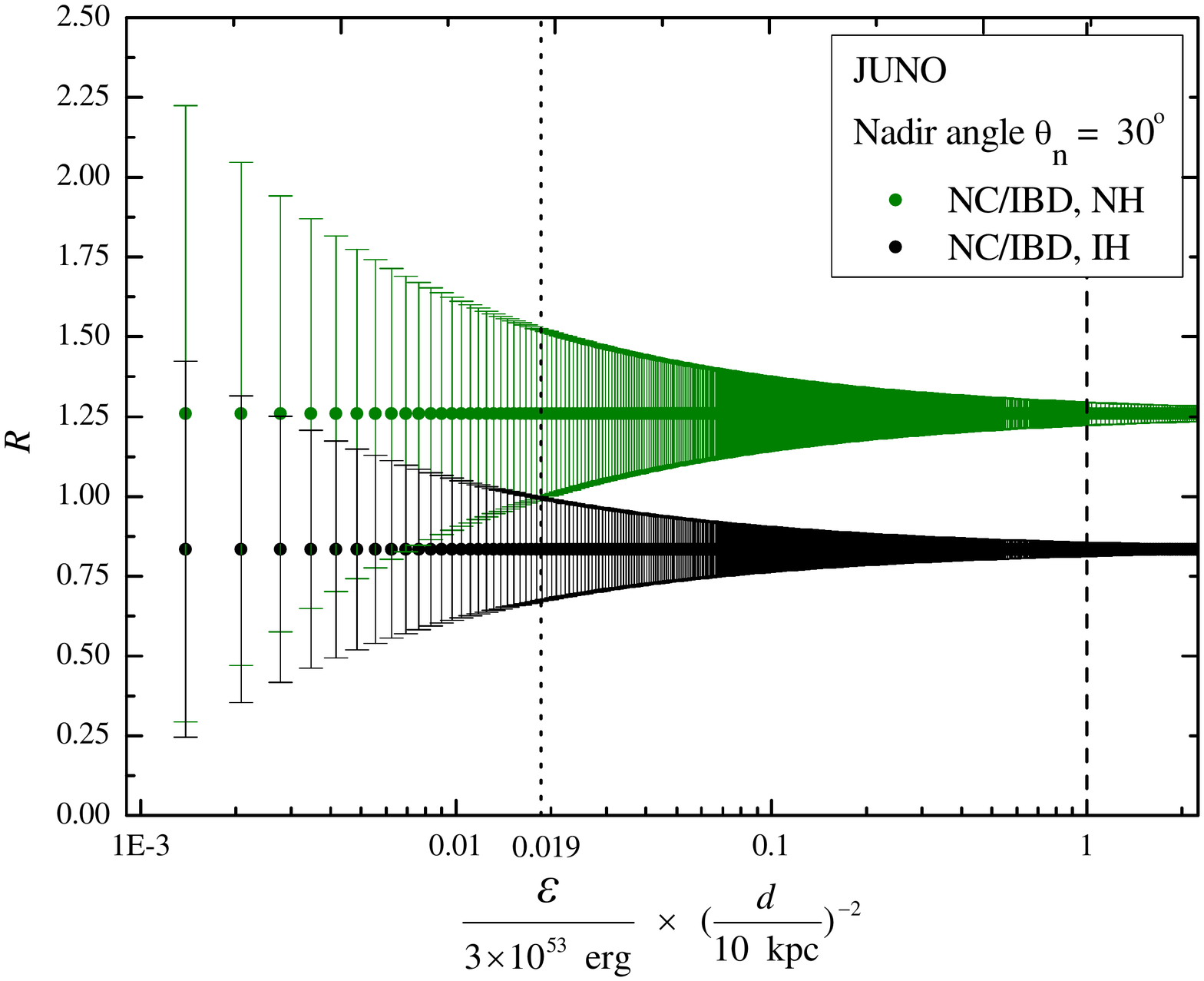}
	\includegraphics[width=8cm]{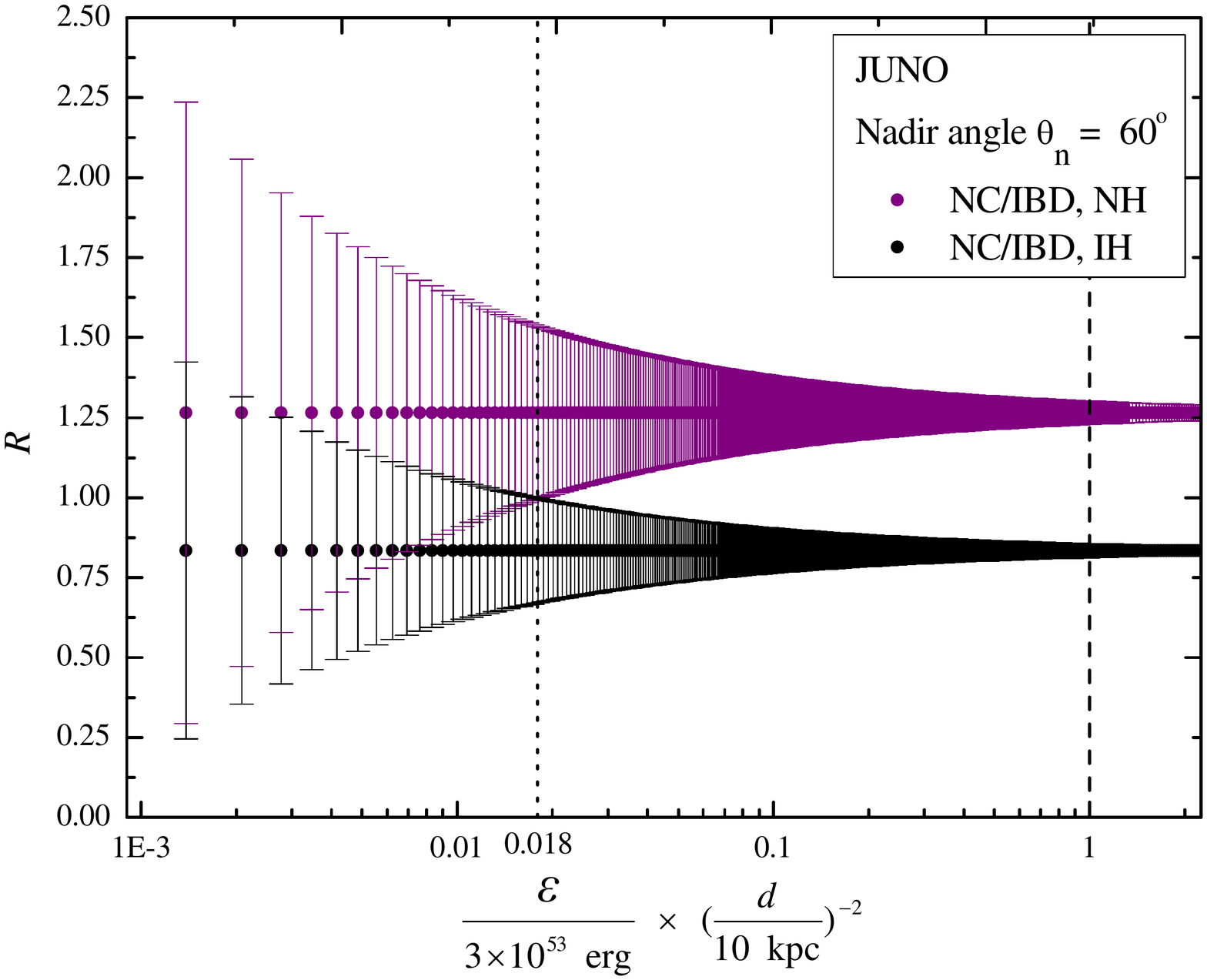}
	\includegraphics[width=8cm]{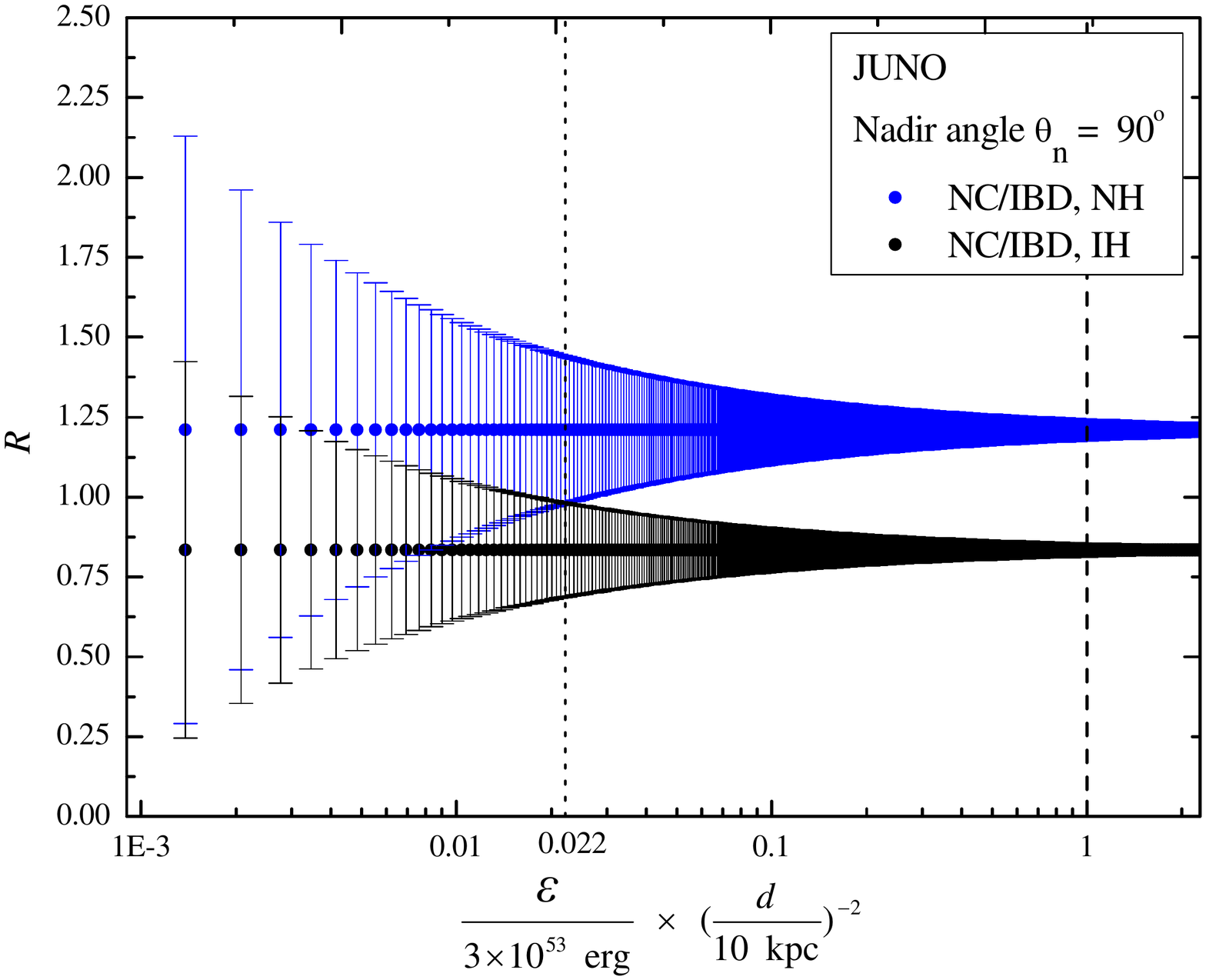}
	\caption{Ranges of $R$ for normal and inverted hierarchies expected at JUNO detector}
	\label{fig:projectedRJuno}
	\end{center}
\end{figure}

The event number in Borexino is smallest hence the statistical uncertainty $\sigma_R$ of this detector is largest. This leads to the overlapping in $R$ for normal and inverted mass hierarchies. Meanwhile, a less energetic SN at a more distant site will also yield fewer events, diminishing the detector capability to resolve the neutrino mass hierarchy. The number of events depends not only on the size of the detector but also on the energy and location of the supernova.
In terms of the total energy of the supernova and its distance to the Earth, we plot the ranges of $R$ for KamLAND detector in Fig. \ref{fig:projectedR}. In our approach, the capability for a specific detector to resolve neutrino mass hierarchy depends on the value of $k\equiv({\mathcal E}/3\times10^{53}{\rm erg})(d/10{\rm kpc})^{-2}$. In Fig. \ref{fig:projectedR}, $k$ = 1 (dashed line) corresponds to the detection of a typical supernova with ${\mathcal E}=3\times10^{53}~{\rm erg}$ at $d=10~{\rm kpc}$. Larger $k$ represents a closer or more energetic SN. The dotted line indicates the threshold for $k$, below which the detector is not able to statistically distinguish between NH and IH due to low number of interactions. As one can see, when $k$ is greater than $1$, the ranges of $R$ for normal and inverted mass hierarchies do not overlap. In addition, we also plot the ranges $R$ for normal and inverted hierarchies expected at JUNO in Fig. \ref{fig:projectedRJuno}.

\label{reference}

\section{The Effect of Luminosity Evolution}

In the above section, we test our method of identifying the neutrino mass hierarchy with an assumption of energy equipartition between all flavors of neutrinos during the entire SN neutrino burst. In reality, neutrino emissions evolve with time as the SN explodes. Not only the luminosities but also the mean energies of all flavors changes with time resulting in time dependent neutrino fluxes of different flavors. The time window of the SN neutrino burst is of the order of $\sim10$ seconds. The entire duration can be classified into three phases: neutronization, accretion, and cooling phases. Though different in detail, various simulations on core-collapse SNe exhibit similar evolution of luminosities and mean energies for each flavor. Simulations indicate that, during most time of the SN neutrino burst (including the accretion and cooling phases), luminosities of all flavors are comparable, except for the very early time of neutronlization during which the luminosities of $\bar{\nu}_e$ and $\nu_x$ are negligibly small compared with that of $\nu_e$. 

An important feature of the evolution of SN neutrino emissions is that the hierarchy of luminosities in accretion and cooling phases are reversed. Our present understanding is that ${\mathcal L_{\nu_e}}\approx{\mathcal L_{\bar{\nu}_e}}>{\mathcal L_{\nu_x}}$ during the accretion phase and ${\mathcal L_{\nu_e}}\approx{\mathcal L_{\bar{\nu}_e}}<{\mathcal L_{\nu_x}}$ during the cooling phase. The luminosity ratio between different flavors, ${\mathcal L_{\nu_e}}/{\mathcal L_{\nu_x}}$, varies across models. An allowed range for this ratio has been suggested \cite{Lunardini:2003eh} as $0.5\lesssim{\mathcal L_{\nu_e}}/{\mathcal L_{\nu_x}}\lesssim2$.  The neutrino emission during neutronlization is small compared with the total emission during the entire burst. Hence we neglect this phase and take a two-phase scenario for modelling the time evolution of the SN neutrino emission. We shall calculate the ratio $R$ for each phase. For each of the accretion phase and cooling phase, we choose fixed ${\mathcal L_{\nu_e}}/{\mathcal L_{\bar{\nu}_e}}$ and ${\mathcal L_{\nu_e}}/{\mathcal L_{\nu_x}}$. To make comparison with the reference equipartition scenario, we also assume $<E_{\nu_e}>=12~{\rm MeV}$, $<E_{\bar{\nu}_e}>=15~{\rm MeV}$, and $<E_{\nu_x}>=18~{\rm MeV}$ for both accretion phase and cooling phase and the same total energy output ${\mathcal E}=3\times10^{53}~{\rm erg}$. Referring to the SN simulations (see \cite{Mirizzi:2015eza} and references therein), we choose the energy ratios between flavors in each phase as 
\begin{equation}
{\mathcal E}_{\nu_e,{\mathcal A}}:{\mathcal E}_{\bar{\nu}_e,{\mathcal A}}:{\mathcal E}_{\nu_x,{\mathcal A}}:{\mathcal E}_{\nu_e,{\mathcal  C}}:{\mathcal E}_{\bar{\nu}_e,{\mathcal  C}}:{\mathcal E}_{\nu_x,{\mathcal  C}} = 30:30:24:22:22:25,
\end{equation}
where ${\mathcal A}$ and ${\mathcal  C}$ denote the accretion and cooling phases, respectively. Also, we denote the entire duration of the two-phase scenario by ${\mathcal  D}$ and the reference equipartition scenario, which has been discussed in Sec. \ref{reference}, by ${\mathcal  Q}$.

\begin{table}[htbp]
\begin{center}
\begin{tabular}{llrrrcccc}\hline\hline
               &   & \multicolumn{1}{c}{NC}  & \multicolumn{2}{c}{IBD} & \multicolumn{2}{c}{R} & \multicolumn{2}{c}{$\sigma_{\rm R}[10^{-2}]$} \\ \cline{4-9}
              &       &        &  IH    & NH$~90^\circ$   &   IH   &   NH$~90^\circ$   &   IH     & NH$~90^\circ$ \\ \hline
                    & Accretion   &   70    &  81   &  63       & 0.87  & 1.11   & 18.0  & 23.8     \\ 
KamLAND    & Cooling     &  69     & 84    &  54       & 0.82  & 1.27  & 16.9  & 28.2     \\
                  & Throughout  &  139   & 165  & 117      & 0.84  & 1.18  & 12.3  & 18.3         \\  \hline
                  & Accretion     &  1704  & 1964  & 1541  & 0.87  & 1.11  & 3.25  & 4.34      \\
JUNO        & Cooling        &  1673  & 2046  & 1317  & 0.82  & 1.27  & 3.06  & 5.19        \\ 
                  & Throughout  &  3377  & 4010  & 2858  & 0.84  & 1.18  & 2.23  & 3.34     \\ \hline

\end{tabular}
\end{center}
\caption{Numbers of NC and IBD interactions in different phases of the SN neutrino burst} 
\label{twophase}
\end{table}

\begin{figure}[htbp]
	\begin{center}
	\includegraphics[width=8cm]{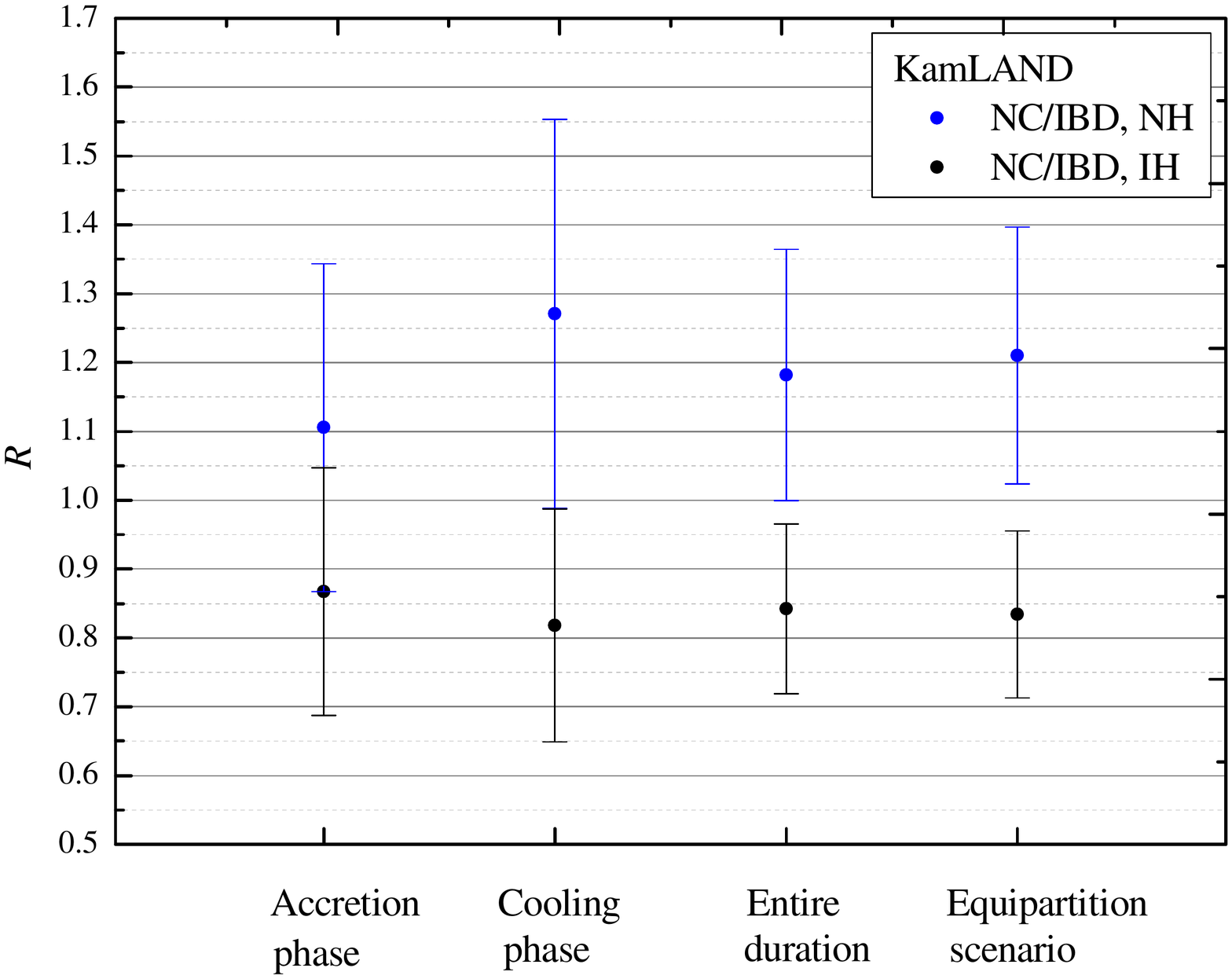}
	\includegraphics[width=8cm]{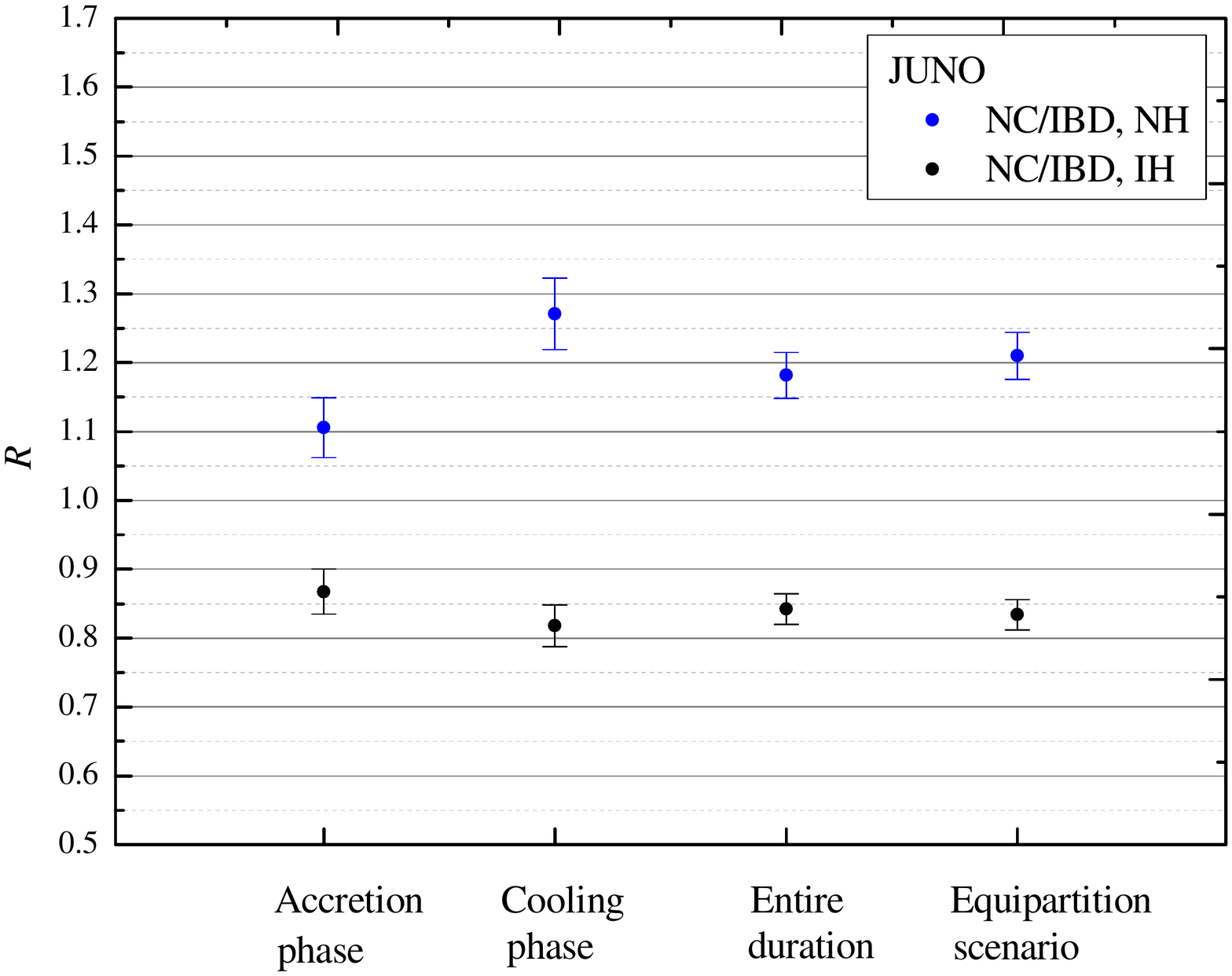}
	\caption{Expected ranges of $R$ for accretion and cooling phases at KamLAND and JUNO detectors in a two-phase scenario of the SN neutrino burst. In the normal hierarchy, we only present the case for no earth-crossing, $\theta_{\rm n}=90^\circ$, for simplicity.}
	\label{fig:twophase}
	\end{center}
\end{figure}

The ranges of $R$ for ${\mathcal A}$, ${\mathcal C}$, ${\mathcal  D}$, and ${\mathcal Q}$ are presented in Fig. \ref{fig:twophase} and the $R$ values are shown in Table \ref{twophase}. It is shown that the ranges of $R$ for ${\mathcal A}$ and ${\mathcal C}$ become larger than those for ${\mathcal D}$ and ${\mathcal Q}$ due to smaller total fluences, resulting from reduced total energy outputs. The ranges of $R$ for ${\mathcal D}$ and ${\mathcal Q}$ are comparable because the same energy output produces similar numbers of neutrinos although the energy partition in the two cases are different. For KamLAND, the neutrino fluence in each of ${\mathcal A}$ and ${\mathcal C}$ phase is not enough to distinguish between neutrino mass hierarchies using our method with a typical SN. A clear discrimination requires observations over the entire duration of the SN neutrino burst. JUNO is large enough to collect more events. Hence, in each of ${\mathcal A}$ and ${\mathcal C}$ phase, neutrino mass hierarchies can be discriminated.

In the two-phase scenario, $R$ values in ${\mathcal A}$ and ${\mathcal C}$ are not only different from each other but also different from those in ${\mathcal D}$ and ${\mathcal Q}$. The difference arises from the deviation from energy equipartition and consequently the changes in neutrino fluence proportion for each flavor. As the neutrino burst evolves from ${\mathcal A}$ to ${\mathcal C}$, the luminosity ratio is reversed from ${\mathcal L_{\bar{\nu}_e}}/{\mathcal L_{\nu_x}}>1$ to ${\mathcal L_{\bar{\nu}_e}}/{\mathcal L_{\nu_x}}<1$. Therefore, $R$ changes from smaller to larger than those in ${\mathcal D}$ and ${\mathcal Q}$ for the normal hierarchy. For the inverted hierarchy, $R$ changes inversely since the matter effect inside the SN swaps $\bar{\nu}_e$ and $\nu_x$.

Although $R$ changes as the luminosity of each flavor evolves, Fig. \ref{fig:twophase} shows that the neutrino mass hierarchy can still be resolved with the two-phase scenario. The result obtained in this simplified model indicates that the evolution of the mean energy of each flavor can also result in different $R$ values since the latter is sensitive to the energy spectrum for each flavor. A more detailed study is beyond the scope of this paper and shall be left for the future publication.

\section{Summary and Conclusions}

We have presented a method for identifying the neutrino mass hierarchy by detecting SN neutrinos with scintillation detectors. In our approach, IBD events of electron anti-neutrinos and NC events of all flavors are considered to derive interaction spectra of IBD and NC. Results on flavor-dependent SN neutrino fluences arriving at Earth indicate that the ratio $R$ of total NC interactions to IBD ones above the selected energy cut $E_{\nu,{\rm s}}$ should be larger than one for the normal mass hierarchy and smaller than one for the inverted mass hierarchy. We not only calculate the expected value of $R$ but also take into account statistical fluctuations of $R$ arising from measurements. This allows one to check the detector capability for resolving the neutrino mass hierarchy.

We have tested our approach with five detectors of different scales and found that $R\cong0.84$ for the inverted hierarchy and $R\cong1.25$ for the normal hierarchy. With neutrinos from a typical SN, the ranges of $R$ from two mass hierarchies do not overlap in all considered detectors except Borexino which has the smallest number of target protons. Besides the number of free protons, the quenching behavior of the liquid also affects the number of neutrino events. This explains the difference on event numbers between KamLAND and SNO+ even though both detectors have almost the same number of target protons. In addition, we have also presented the detector capability for mass hierarchy discrimination for different total energy outputs and locations of SNe, as shown in Figs. \ref{fig:projectedR} and \ref{fig:projectedRJuno}.

Earth matter effect on resolving the neutrino mass hierarchy was studied by considering $\bar{\nu}_e$ fluences at different nadir angles. We found that the Earth matter effect has only negligible influence on $R$. However, with a large number of events, our method could be applied to investigate the Earth matter effect by considering the ratio of $(dN/dE)_{\rm NC}$ to $(dN/dE)_{\rm IBD}$, which is energy-dependent. In this work, we assume MSW effect is the dominant mechanism for the flavor transition inside SN. The total number of NC interactions is not only a useful normalization for defining $R$ but also crucial for determining the total neutrino fluence. 

We have also tested our method with respect to a simplified two-phase model for the evolution of SN neutrino emission. We found that $R$ changes as the relative luminosity of each neutrino flavor changes from the accretion phase to the cooling phase during the SN neutrino evolution. Further studies with more realistic models will be addressed in the future works together with the inclusion of collective effect on flavor transition of SN neutrinos.

\section*{Acknowledgements}
We thank J. Beacom, B. Dasgupta, S.-H. Chiu and L. Wen for helpful discussions and comments. This work is partly supported by the Ministry of Science and Technology, Taiwan, under Grants No. MOST 104-2112-M-009-021 and MOST 104-2112-M-009-006.

\end{document}